\documentclass[a4paper,11pt]{article}
\pdfoutput=1
\usepackage{jheppub}
\usepackage[T1]{fontenc}
\usepackage{graphicx}
\usepackage{epsfig}
\usepackage{rotating}
\usepackage{amssymb}
\usepackage{subfigure}
\usepackage{dsfont}
\usepackage{psfrag}
\usepackage{amsmath,euscript,array,mathrsfs}
\usepackage{axodraw}
\usepackage{bbold}
\usepackage{epsf}

\newcommand{\beq}{\begin{equation}}
\newcommand{\eeq}{\end{equation}}
\newcommand{\beqs}{\begin{eqnarray}}
\newcommand{\eeqs}{\end{eqnarray}}

\newcommand{\Tr}{{\rm Tr}}

\def\hbar{\hspace{0pt}\raisebox{1pt}{$-$} \hspace{-7pt} h}

\def\di{\mbox{d}}
\def\r{\rho}

\newcommand{\be}{\begin{equation}}
\newcommand{\ee}{\end{equation}}
\newcommand{\bea}{\begin{eqnarray}}
\newcommand{\eea}{\end{eqnarray}}

\def\lbldef#1#2{\expandafter\gdef\csname #1\endcsname {#2}}

\def\href#1#2{#2}
\def\half{{1 \over 2}}

\newcommand{\ber}{\begin{eqnarray}}
\newcommand{\eer}{\end{eqnarray}}

\newcommand{\beqar}{\begin{eqnarray}}

\newcommand{\eeqar}{\end{eqnarray}}

\newcommand{\dsl}{\kern.06em\hbox{\raise.15ex\hbox{$/$}\kern-.56em\hbox{$\partial$}}}

\newcommand{\eeqarr}{\end{eqnarray}}
\newcommand{\ZZ}{{\rm \kern 0.275em Z \kern -0.92em Z}\;}

\def\CC{{\mathchoice
{\rm C\mkern-8mu\vrule height1.45ex depth-.05ex
width.05em\mkern9mu\kern-.05em}
{\rm C\mkern-8mu\vrule height1.45ex depth-.05ex
width.05em\mkern9mu\kern-.05em}
{\rm C\mkern-8mu\vrule height1ex depth-.07ex
width.035em\mkern9mu\kern-.035em}
{\rm C\mkern-8mu\vrule height.65ex depth-.1ex
width.025em\mkern8mu\kern-.025em}}}

\def\RR{{\rm I\kern-1.6pt {\rm R}}}

\def\ZZ{{\rm Z}\kern-3.8pt {\rm Z} \kern2pt}
\def\IB{\relax{\rm I\kern-.18em B}}
\def\ID{\relax{\rm I\kern-.18em D}}
\def\II{\relax{\rm I\kern-.18em I}}
\def\IP{\relax{\rm I\kern-.18em P}}

\newcommand{\bear}{\begin{eqnarray}}
\newcommand{\eear}{\end{eqnarray}}

\def\r{\rho}

\def\tt{\tilde{\theta}}

\def\6{\partial}

\def\bea{\begin{eqnarray}}
\def\eea{\end{eqnarray}}

\def\beqx{\begin{displaymath}}
\def\eeqx{\end{displaymath}}

\newcommand{\bmat}{\left(\begin{array}}
\newcommand{\emat}{\end{array}\right)}

\def\half{\frac{1}{2}}

\def\r{\rho}

\def\bo{{\raise-.3ex\hbox{\large$\Box$}}}               
\def\face{{\raise.2ex\hbox{$\displaystyle \bigodot$}\mskip-2.2mu \llap {$\ddot
        \smile$}}}                                   
\def\>{\rangle}                                      
\def\<{\langle}                                      

\def\leftrightarrowfill{$\mathsurround=0pt \mathord\leftarrow \mkern-6mu
        \cleaders\hbox{$\mkern-2mu \mathord- \mkern-2mu$}\hfill
        \mkern-6mu \mathord\rightarrow$}        
\def\dvec#1{\vbox{\ialign{##\crcr
        \leftrightarrowfill\crcr\noalign{\kern-1pt\nointerlineskip}
        $\hfil\displaystyle{#1}\hfil$\crcr}}}           
\def\Tr{{\rm Tr \,}}                                    

\def\-{\hphantom{-}}
\newcommand{\dd}{\mbox{d}}

\title{Holographic glueballs from the circle reduction of Romans supergravity}

\author[a,b]{Daniel Elander,}
\author[c]{Maurizio Piai,}
\author[c]{John Roughley}

\affiliation[a]{Laboratoire Charles Coulomb (L2C), University of Montpellier, CNRS, Montpellier, France}
\affiliation[b]{Departament de F\'isica Qu\`antica i Astrof\'isica \& Institut de Ci\`encies del Cosmos (ICC), Universitat de Barcelona, Mart\'i Franqu\`es 1, ES-08028, Barcelona, Spain}
\affiliation[c]{Department of Physics, College of Science, Swansea University,
Singleton Park, SA2 8PP, Swansea, Wales, UK}

\date{\today}

\abstract{
We reconsider a one-parameter class of known solutions of the circle compactification of Romans six-dimensional half-maximal supergravity. The gauge-theory duals of these solutions are confining four-dimensional field theories. Their UV completions consist of the compactification on a circle of a higher-dimensional field theory that is flowing between two fixed points in five dimensions. We systematically study the bosonic fluctuations of the supergravity theory, corresponding to the bosonic glueballs of the dual field theory.

We perform numerically the calculation of the spectrum of excitations of all the bosonic fields, several of which had been disregarded in earlier work on the subject. We discuss the results as a function of the one parameter characterising the class of background solutions, hence further extending known results. We show how certain towers of states are independent of the background, and compare these states to existing lattice literature on four-dimensional Yang-Mills (pure) gauge theories, confirming the existence of close similarities.

For the aforementioned analysis, we construct gauge-invariant combinations of the fields appearing in the 
reduction to five dimensions of the supergravity theory, and hence focus on the 32 physical bosonic degrees of freedom. We show explicitly how to implement gauge-fixing of the supergravity theory. The results of such technical work could be used to analyse the spectra of other theories proposed in the context of top-down holography. For example, it could be applied to holographic realisations of composite-Higgs and light-dilaton scenarios.
}

\begin{document}
\maketitle

\section{Introduction}

The study of strongly-coupled, confining theories in four dimensions is notoriously difficult. Understanding the non-perturbative dynamics of these theories is of vital importance for particle physics, not only because Quantum Chromo-Dynamics (QCD) is one such theory, but also because many elegant solutions to the hierarchy problem(s) of the electro-weak (EW) theory rely on the existence of new strongly-coupled dynamics. The long distance behaviour of realistic EW models cannot resemble that of QCD, as was the case in traditional Technicolor models, that have been excluded by experimental data. Examples of phenomenologically viable proposals yield, at low energy, either a light dilaton or a set of composite Higgs fields that originate dynamically  as pseudo-Goldstone bosons. It is hence desirable to study observable quantities that provide information about the underlying dynamics and might be used to select theories with interesting phenomenology.

The spectrum of physical particles in Yang-Mills theories in $D=4$ dimensions consists of discrete, gauge-invariant bound states, the glueballs, with typical mass ${\cal O}(\Lambda)$, where $\Lambda$ is the scale that is dynamically generated. These  particles appear in correlation functions of gauge-invariant local operators ${\cal O}$ built of gluon fields of the theory, e.g.
\beqs
{\cal O}&\equiv& \Tr F_{\mu\nu}F^{\mu\nu}\,.
\eeqs
On quite general grounds, in a confining theory one expects infinite numbers of such glueballs, that can be classified by quantum numbers of the Poincar\'e group (mass and spin), possibly supplemented by  additional, model-dependent quantum numbers. The literature on the subject includes for instance the reviews in~\cite{Teper:1998kw,Lucini:2012gg}, and the detailed lattice studies in~\cite{Morningstar:1999rf,Chen:2005mg,Lucini:2001ej,Lucini:2004my,Meyer:2004jc,Lucini:2010nv,Athenodorou:2015nba,Lau:2017aom,Bennett:2017kga}, besides suggestive ideas on general properties of the glueball spectra (the literature on which is vast, and deserves being reviewed elsewhere, but see for instance~\cite{Abreu:2005uw,Mathieu:2005wc,Mathieu:2008me,Buisseret:2011bg,Bochicchio:2013eda,Athenodorou:2016ndx,Hong:2017suj}). In more general theories, in which gluons couple to matter fields, the physical particles result from mixing between operators made purely of glue and other operators with the same quantum numbers. With some abuse of language, we still refer to such particles as glueballs.

The study of strongly coupled field theories received a major boost with the advent of gauge-gravity dualities~\cite{Maldacena:1997re,Gubser:1998bc,Witten:1998qj} (for a pedagogical introduction see also~\cite{Aharony:1999ti}). Soon after the earliest studies provided support for the existence of a non-perturbative, weak-strong duality between some special conformal field theories and higher-dimensional gravity theories, it was also proposed that one can extend the duality to gravity models 
that provide the dual description of confining field theories. Most importantly for our purposes, the dictionary governing the calculations of the holographically renormalised 2-point functions (see for instance~\cite{Skenderis:2002wp,Papadimitriou:2004ap}) of relevance to glueball spectra has been established.

Broadly speaking, there are two classes of realisations of such proposal for the dual of four-dimensional confining theories, depending on the geometric realisation of confinement. Along the original suggestion in~\cite{Witten:1998zw}, by toroidal compactification of supergravities admitting AdS$_D$ backgrounds one may be able to find smooth solutions in which one of the internal circles shrinks to zero size at a finite value of the radial direction. The spectrum of glueballs in this case resembles qualitatively what is expected in the case of QCD-like theories (see for instance~\cite{Brower:2000rp}, and references therein for earlier attempts). In particular, there are no known examples of this type in which one of the four-dimensional scalar particles becomes anomalously light, in contrast to what is expected in the presence of dilaton dynamics. Yet, one must wonder  whether such models can be used as the dynamical origin of more general composite-Higgs models. Addressing this possibility requires computing the spectrum of the whole physical sector captured by supergravity, including 0-forms, 1-forms and 2-forms in the bosonic sector.

For completeness, we remind the reader of a second class of supergravity backgrounds modelling the dual of confining gauge theories, that is related to the deformation and resolution of the conifold~\cite{Candelas:1989js,Klebanov:1998hh,Klebanov:2000nc,PandoZayas:2000ctr}, and includes for example Refs.~\cite{Klebanov:2000hb,Maldacena:2000yy,Chamseddine:1997nm,Butti:2004pk}. In these backgrounds, the geometry is characterised by the fact that a $2$-sphere shrinks at the end of space in the radial direction. Variations of these backgrounds show that the scalar glueballs may include a parametrically light particle~\cite{Nunez:2008wi,Elander:2009pk,Elander:2012yh,Elander:2014ola}. A non-trivial example of this has been identified within a five-dimensional sigma-model system, the background solutions of which are lifted to $D=10$ dimensions to provide the gravity dual of the baryonic branch of the Klebanov-Strassler system~\cite{Elander:2017hyr,Elander:2017cle}. It would be interesting to compute the spectrum also of other modes, besides the scalars appearing in the Papadopoulos-Tseytlin ansatz~\cite{Papadopoulos:2000gj}, in all these backgrounds, in order to understand the structure of global symmetries (and supersymmetries) in detail, for example by considering the consistent truncation in~\cite{Cassani:2010na,Bena:2010pr}. 

In order to perform both tasks---namely to start studying composite-Higgs models in the rigorous top-down holographic approach,
but also characterising the full set of symmetries of the existing supergravity models yielding dilaton dynamics at low energies---one must explicitly  keep track of gauge invariance  in the calculations performed in the five-dimensional theories with boundaries. In particular, only gauge-invariant modes belong in the  physical spectra. There are a number of subtleties involved in doing so, and with this paper we contribute to the programme of systematic explorations of the bosonic spectra of gravity theories dual to confining gauge theories in $D=4$ dimensions, by first considering one of the simplest of the models of the first class: the smooth supergravity backgrounds in $D=5$ dimensions obtained by reduction on a circle of the $F_4$ gauged supergravity theory in $D=6$ dimensions~\cite{Romans:1985tw}. In the future, we envision applying the process developed in this paper to other more complicated supergravity theories. In particular, it would be interesting to consider Witten's model~\cite{Witten:1998zw} and its extension in~\cite{Elander:2013jqa}, by performing a parallel study of the complete supergravity theory in $D=7$ dimensions it belongs to, hence extending the results of~\cite{Brower:2000rp}. 

Extended supergravities admitting supersymmetric anti-de-Sitter solutions in $D$ space-time dimensions have been classified by Nahm~\cite{Nahm:1977tg} (see also~\cite{Kac:1977em}). A special case is the ${\cal N}=(2,2)$, non-chiral, half-maximal ($16$ supercharges), gauged  supergravity  in $D=6$ dimensions with gauge group $SU(2)$ and $F_4$ superalgebra, predicted in~\cite{DeWitt:1981wm} and constructed by Romans in~\cite{Romans:1985tw}. It can be obtained from massive type-IIA in $D=10$ dimensions~\cite{Romans:1985tz}, via a consistent warped $S^4$ reduction that preserves  an  $SO(4)$ symmetry of the internal space, and breaks half of the supersymmetry~\cite{Brandhuber:1999np,Cvetic:1999un}.\footnote{Alternative embeddings in Type IIB involve an internal space with less symmetry \cite{Jeong:2013jfc,Hong:2018amk}.} One of the angles parametrising the internal manifold enters non-trivially into the expression of the warp factor in the lift from $6$ to $10$ dimensions, which vanishes at the equator, so that  the internal geometry is in fact a foliation of $3$-spheres, broadly corresponding to the upper hemisphere of $S^4$. We refer the reader to the literature for details that do not play a central role in this paper. 

The scalar manifold of the $D=6$, half-maximal, non-chiral theories is described by one of the following cosets~\cite{DAuria:2000afl,Andrianopoli:2001rs} (see also~\cite{Freedman:2012zz,Tanii:2014gaa}):
\beqs
\frac{O(4,n)}{O(n)\times SO(4)}\,\times \,O(1,1)\,,
\eeqs
where the pure, non-chiral supergravity theory is coupled to $n$ vector multiplets, each of which contains a vector field, four spin-$\frac{1}{2}$ fields and four real scalar fields.\footnote{In counting fermionic degrees of freedom, we follow these conventions:
because the symplectic Majorana condition and the chirality condition in $D=6$ dimensions can be imposed simultaneously, a single Dirac fermion consisting of $2^{D/2}=8$ complex spinorial components can be decomposed in the sum of $2$ left-handed and $2$ right-handed symplectic Majorana-Weyl spinors (making the $SU(2)\times SU(2)\sim SO(4)$ symmetry manifest) giving a total of four symplectic Majorana-Weyl spinors, each of which can be written as a quaternionic field. Hence the vector multiplet 
contains $8$ bosonic and $8$ fermionic degrees of freedom. We refer to these theories as ${\cal N}=(2,2)$, because each of the four supersymmetries is generated  by a symplectic Majorana-Weyl spinor, which is represented by a quaternion, or equivalently by 4 real components, for a total of $4\times 4=16$ supercharges. It might be useful to the reader to notice that half of the supergravity literature refers to this same theory, with the same amount of supersymmetry, as ${\cal N}=(1,1)$.} The compact $SO(4)\sim SU(2)\times SU(2)$ global symmetry contains the diagonal $SU(2)$ $R$-symmetry. Such theories have attracted some interest in the context of the AdS$_6$/CFT$_5$ correspondence (see for instance~\cite{Nishimura:2000wj,Ferrara:1998gv}), of the holographic study of non-trivial renormalisation group flows (see for instance~\cite{Gursoy:2002tx,Nunez:2001pt,Karndumri:2012vh}) and of non-abelian T-duality (see for instance~\cite{Lozano:2012au,Jeong:2013jfc}).

In this paper, we restrict attention to the $n=0$ pure supergravity case, in which the scalar manifold reduces to $ \mathbb{R}_+$,
and is parameterised by the scalar $\phi$. The field content consists of the supergravity multiplet: the graviton (propagating $(D-1)(D-2)/2-1=9$ degrees of freedom on-shell), one 2-form ($(D-2)(D-3)/2=6$ degrees of freedom), four vectors ($4\times (D-2)=16$ degrees of freedom), one real scalar, four symplectic Majorana-Weyl gravitini ($4\times(D-3)2^{D/2}/4=24$ degrees of freedom) and four symplectic Majorana-Weyl spin-$\frac{1}{2}$ fields ($4\times 2^{D/2}/4=8$ degrees of freedom). This theory admits two distinct critical points, only one of which preserves supersymmetry, though both are perturbatively stable.
 
By compactifying one dimension on a circle one deforms the  AdS$_6$ solutions in such a way as to realise a simple  dual description of a four-dimensional confining theory, along the lines of~\cite{Witten:1998zw}. Several interesting studies of parts of the spectrum of glueballs of the dual field theory have been published before (see in particular~\cite{Wen:2004qh,Kuperstein:2004yf,Elander:2013jqa}), in which the fluctuations of the supergravity backgrounds are computed explicitly.

Following~\cite{Elander:2013jqa}, we consider classical backgrounds in which the solutions for $\phi$ interpolate between the two known, (perturbatively) stable critical points of the $D=6$ theory, while we also compactify one of the space-like coordinates on a shrinking circle. The solutions provide a one-parameter family of backgrounds that at low energy describe confining four-dimensional  dual theories. The one parameter is denoted by $s_{\ast}$ in the following, and it encodes the parametric separation between the scale of confinement in the dual four-dimensional theory and the scale of the flow between the two fixed points in the UV-complete five-dimensional gravity theory. We complete the existing literature by computing the spectrum of all the bosonic modes associated with the fields appearing in the action in $D=6$ dimensions. To do so, we fluctuate all the fields, linearising the resulting equations of motion and introducing appropriate gauge-invariant combinations. We obtain new, previously unknown results, and we show that the bosonic modes may be classified into two distinct groups, characterised by the two very different ways in which the modes behave as a function of $s_{\ast}$. In the process, we elucidate on the subtleties connected with gauge-invariance, that are of general applicability to more complicated systems.

The paper is organised as follows. In Section~\ref{Sec:Model}, we report the six-dimensional action, and perform its reduction on a circle to five dimensions. We also summarise known results about the classical solutions of the theory. In Section~\ref{Sec:Mass}, we report on our calculation of the spectra of fluctuations of all the bosonic (physical) degrees of freedom in the five-dimensional backgrounds of interest. In Section~\ref{Sec:Discussion} we discuss the physical meaning of our results, and compare them to the literature. In Section~\ref{Sec:Outlook} we outline future work for which this paper lays the foundations.

Appendix~\ref{Sec:A} and~\ref{Sec:B} deal, respectively, with general results in four- and five-dimensional bosonic theories.
Appendix~\ref{Sec:A1} contains some useful conventions about the notation and some well known results about the treatment of
massive vectors in $D=4$ dimensions.

Appendix~\ref{Sec:A2} is a somewhat digressive technical section. We find it useful to remind the reader how the equivalence of massive 2-form and massive 1-forms in $D=4$ dimensions can be made manifest, and for this purpose we follow~\cite{Bijnens:1995ii} (see also~\cite{Ecker:1989yg,Bruns:2004tj}), and make explicit the role of the gauge redundancies in the two formulations, in particular in reference to the Higgs mechanism. We also comment briefly on what happens in higher dimensions, on how
the dualities between forms of different order affect the Higgs and the soldering  phenomena (see for instance~\cite{Noronha:2003vp}), and on some of the subtleties emerging in the context of gauged supergravities (see for instance~\cite{Samtleben:2008pe}  and references therein).

Appendix~\ref{Sec:B1} contains a summary of material borrowed from~\cite{Bianchi:2003ug,Berg:2005pd,Berg:2006xy,Elander:2009bm,Elander:2010wd}, that describes and explains the gauge-invariant formalism we adopt in the treatment of scalar and tensor fluctuations of the five-dimensional backgrounds. Appendix~\ref{Sec:B2} and~\ref{Sec:B3} deal with the gauge-fixing of the bulk and boundary actions of 1-forms and 2-forms, respectively.

\section{The model}
\label{Sec:Model}

\subsection{Action and formalism of the six-dimensional model}

As anticipated in the Introduction, our starting point is the  supergravity in $D=6$ dimensions written by Romans in~\cite{Romans:1985tw}, that can also be obtained as warped reduction on $S^{4}$  of the ten-dimensional massive Type-IIA supergravity theory.  
We label six-dimensional quantities by hatted Roman indices as $\hat{M} = 0 , 1 , 2 , 3 , 5 , 6$, and adopt the
convention in which the metric has signature mostly plus. The 32 degrees of freedom of the bosonic part of the six-dimensional action are written in terms of the scalar $\phi$, the metric $\hat g_{\hat{M}\hat{N}}$, a $U(1)$ vector $A_{\hat{M}}$ and its field strength $\hat F_{\hat{M}\hat{N}}$, three vectors $A_{\hat{M}}^i$ transforming on the  adjoint of $SU(2)$ and their field strengths $\hat F_{\hat{M}\hat{N}}^i$ and the 2-form $B_{\hat{M}\hat{N}}$ and its field strength $\hat G_{\hat{M}\hat{N}\hat{T}}$. We omit topological terms  that start at the cubic order in $\hat F_{\hat{M}\hat{N}}$, $B_{\hat{M}\hat{N}}$ and $\hat F_{\hat{M}\hat{N}}^{i}$, as they vanish on the backgrounds of interest and do not enter the (linearised) equations for the fluctuations, so that the action we consider is given by
{\small
\begin{align}
S_{6}=\int_{}^{}\text{d}^{6}x\sqrt{-\hat g_{6}}
\bigg(\frac{\mathcal{R}_{6}}{4}&- \hat g^{\hat{M}\hat{N}}\partial_{\hat{M}}\phi\partial_{\hat{N}}\phi -\mathcal{V}_{6}(\phi)  
-\frac{1}{4}e^{-2\phi}\hat g^{\hat{M}\hat{R}}
\hat g^{\hat{N}\hat{S}}\sum_{i}\hat F_{\hat{M}\hat{N}}^{i}\hat F_{\hat{R}\hat{S}}^{i}\,+ \notag \\
&-\frac{1}{4}e^{-2\phi}\hat g^{\hat{M}\hat{R}}
\hat g^{\hat{N}\hat{S}}\hat{\mathcal{H}}_{\hat{M}\hat{N}}\hat{\mathcal{H}}_{\hat{R}\hat{S}}
-\frac{1}{12}e^{4\phi}\hat g^{\hat{M}\hat{R}}
\hat g^{\hat{N}\hat{S}}\hat g^{\hat{T}\hat{U}}\hat G_{\hat{M}\hat{N}\hat{T}}\hat G_{\hat{R}\hat{S}\hat{U}} \bigg)\,,
\label{ActionS6}
\end{align} }%
with\footnote{Complete anti-symmetrisation is normalised so that $[n_1n_2\cdots n_p]\equiv \frac{1}{p!}(n_1n_2\cdots n_p-n_2n_1\cdots n_p +\cdots)$.}
\begin{align}
\hat F_{\hat{M}\hat{N}}^{i} &\equiv \partial_{\hat{M}}A_{\hat{N}}^{i}-
\partial_{\hat{N}}A_{\hat{M}}^{i}+g\epsilon_{ijk}A_{\hat{M}}^{i}A_{\hat{N}}^{j}\,,\\
\hat F_{\hat{M}\hat{N}} &\equiv \partial_{\hat{M}}A_{\hat{N}}-
\partial_{\hat{N}}A_{\hat{M}}\,,\\
\hat{\mathcal{H}}_{\hat{M}\hat{N}} &\equiv \hat F_{\hat{M}\hat{N}} + mB_{\hat{M}\hat{N}}\,,\\
\hat G_{\hat{M}\hat{N}\hat{T}}&\equiv 3\partial_{\small[\hat{M}}B_{\hat{N}\hat{T}\small]}=
\partial_{\hat{M}}B_{\hat{N}\hat{T}}+
\partial_{\hat{N}}B_{\hat{T}\hat{M}}+
\partial_{\hat{T}}B_{\hat{M}\hat{N}}\,.
\end{align}
The metric has determinant $\hat g_{6}$, $\mathcal{R}_{6}$ is the corresponding Ricci scalar, {\small$\hat{\mathcal{H}}_{\hat{M}\hat{N}}$} 
couples the $U(1)$ vector and 2-form fields, while {\small$\hat G_{\hat{M}\hat{N}\hat{T}}$} is the field strength tensor of the 2-form. We conventionally fix the units so that the gauge coupling is $g=\sqrt{8}$, and the mass parameter is $m=\frac{2\sqrt{2}}{3}$, while the six-dimensional Newton constant is given by $G_6 = \frac{1}{4\pi}$. The potential for the scalar $\phi$ is
\be
\mathcal{V}_{6}(\phi)=\frac{1}{9}(e^{-6\phi}-9e^{2\phi}-12e^{-2\phi})\,.
\ee
As we shall see, the six-dimensional potential admits two critical points, a maximum and a minimum, and there exist solutions that interpolate between the two.

\subsection{Reduction from $D=6$ to $D=5$ dimensions}

We compactify one of the external dimensions on a circle and look at the resulting five-dimensional system; the size of this circle is parameterised by a new dynamical scalar field $\chi$ that appears in the reduced five-dimensional model. We make use of the following ansatz for the six-dimensional metric:
\be
\text{d}s_{6}^{2}=e^{-2\chi} \text{d}s_{5}^{2} + e^{6\chi}
\big(\text{d}\eta + V_{M}\text{d}x^{M} \big)^{2}\,,
\ee
where $V_{M}$ is naturally defined as covariant, the five-dimensional index is denoted by $M=0,\,1,\,2,\,3,\,5$, the sixth (compact) coordinate is denoted by $\eta$, and we decompose the $SU(2)$ vector fields as $A_{\hat{M}}^{i}=\{A_{\mu}^{i}, A_{5}^{i}, \pi^{i}\}$, where $\mu=0,\,1,\,2,\,3$ is the four-dimensional index.

Compactifying on the circle, according to $\partial_{6}A_{N} = 0 = \partial_{6}B_{NT}$, hence retaining only the zero modes, we find that the action---by ignoring at first the  $U(1)$ fields  $A_{\hat{M}}$ and $B_{\hat{M}\hat{N}}$, i.e. by omitting the last two terms in Eq.~(\ref{ActionS6})---can be rewritten as
\be
\label{eq:S6withtotalder}
S_{6}=\int_{}^{}\text{d}\eta
\bigg\{\tilde S_{5}+\half \int_{}^{}\text{d}^{5}x\,
\partial_{M}\big(\sqrt{-g_{5}}\,g^{MN}\partial_{N}\chi\big)\bigg\} + \cdots \,,
\ee
with the five-dimensional action given by
{\small\be
\tilde S_{5}=\int_{}^{}\text{d}^{5}x\sqrt{-g_{5}}
\bigg(\frac{\mathcal{R}_{5}}{4}-\half G_{ab} g^{MN}\partial_{M}\Phi^{a}\partial_{N}\Phi^{b} -\mathcal{V}(\phi,\chi)  
-\frac{1}{4}H_{AB}g^{MR}g^{NS}F_{MN}^{A}F_{RS}^{B} \bigg)\,.
\ee}%
In this reduced model, the sigma-model scalars are $\Phi^{a} = \{\phi, \chi, \pi^{i}\}$, the potential in $D=5$ dimensions is
\beq
	\mathcal{V}(\phi,\chi) = e^{-2\chi}\mathcal{V}_{6}(\phi) \,,
\eeq
and the metric tensors for the sigma-model scalars as well as the field strengths $\{F^{V},F^{i}\}$ are given by
\beqs
	G_{ab} &=& \text{diag} \Big(2, 6, e^{-6\chi-2\phi}\Big) \,, \\
	 H_{AB} &=& \text{diag}\Big(\frac{1}{4}e^{8\chi},e^{2\chi-2\phi}\Big) \,,
\eeqs
while the field strengths are defined by
\begin{align}
F_{MN}^{V}&\equiv \partial_{M}V_{N}-\partial_{N}V_{M}\,,\\
F_{MN}^{i}&\equiv \partial_{M}A_{N}^{i}-\partial_{N}A_{M}^{i}
+g\epsilon_{ijk}A_{M}^{i}A_{N}^{j}
+(V_{M}\partial_{N}\pi^{i}-V_{N}\partial_{M}\pi^{i})\,.
\end{align} 
The last two terms of Eq.~(\ref{ActionS6}) may be rewritten as follows:
{\small\begin{align}
S_{6}^{U(1)}&=\int_{}^{}\text{d}\eta\text{d}^{5}x\sqrt{-g_{5}}\bigg\{
-\frac{1}{4}H^{(2)}g^{MR}g^{NS}\mathcal{H}_{MN}\mathcal{H}_{RS}
-\frac{1}{12}K^{(2)}g^{MR}g^{NS}g^{TU}G_{MNT}G_{RSU}\,+\notag\\
&\hspace{4cm}-\frac{1}{2}G^{(1)}g^{NS}\mathcal{H}_{6N}\mathcal{H}_{6S}
-\frac{1}{4}H^{(1)}g^{NS}g^{TU}G_{6NT}G_{6SU}
\bigg\}\,,
\end{align}}%
where $H^{(2)}=e^{2\chi-2\phi}$, $K^{(2)}=e^{4\chi+4\phi}$, $G^{(1)}=e^{-6\chi-2\phi}$, $H^{(1)}=e^{-4\chi+4\phi}$, and the decomposition of the tensors in five-dimensional language is governed by the definitions:
\begin{align}
\mathcal{H}_{MN}& \equiv \hat F_{MN}+mB_{MN} + \left( V_M \partial_N A_6 - V_N \partial_M A_6 \right) + m \left( B_{6M} V_N - B_{6N} V_M \right)\,,\\
\mathcal{H}_{6N}& \equiv \hat{\mathcal{H}}_{6N} = \partial_{6}A_{N}-\partial_{N}A_{6}+mB_{6N}=-\partial_{N}A_{6}+mB_{6N}\,,\\
G_{MNT}& \equiv 3\partial_{\small[M}B_{NT\small]}  - 6V_{\small[M} \partial_N B_{T\small]6}\,,\\
G_{6NT}&\equiv \hat G_{6NT} =
\partial_{6}B_{NT}-
\partial_{N}B_{6T}+
\partial_{T}B_{6N}=\partial_{T}B_{6N}-\partial_{N}B_{6T}\,.
\end{align}
The total derivative term in Eq.~\eqref{eq:S6withtotalder} does not affect the equations of motion, and hence we disregard it, so that the complete five-dimensional action we adopt is
{\small\begin{align}
S_{5}=\int_{}^{}\text{d}^{5}x\sqrt{-g_{5}}
\bigg(\frac{\mathcal{R}_{5}}{4}-\half G_{ab} g^{MN}\partial_{M}\Phi^{a}\partial_{N}\Phi^{b} &-\mathcal{V}(\phi,\chi)  
-\frac{1}{4}H_{AB}g^{MR}g^{NS}F_{MN}^{A}F_{RS}^{B}\,\notag\\
-\frac{1}{4}e^{2\chi-2\phi}g^{MR}g^{NS}\mathcal{H}_{MN}\mathcal{H}_{RS}
&-\frac{1}{12}e^{4\chi+4\phi}g^{MR}g^{NS}g^{TU}G_{MNT}G_{RSU}\,\notag\\
-\frac{1}{2}e^{-6\chi-2\phi}g^{NS}\mathcal{H}_{6N}\mathcal{H}_{6S}
&-\frac{1}{4}e^{-4\chi+4\phi}g^{NS}g^{TU}G_{6NT}G_{6SU}
\bigg)\,.\label{Eq:S5ast}
\end{align}}%
The $32$ bosonic degrees of freedom are now described in the five-dimensional action in terms of 6 scalar fields, 6 vector fields ($3$ d.o.f. each), one 2-form field ($3$ d.o.f.), and the metric ($5$ d.o.f.).

\subsection{Classical background solutions}

We write the ansatz for the five-dimensional metric as
\beqs
\di s^2 &=& e^{2A} \di x_{1,3}^2 + \di r^2\,,
\eeqs
with the convention that the four-dimensional metric is  $\eta_{\mu\nu}={\rm diag}\,\left(-\,,\,+\,,\,+\,,\,+\right)$. The radial direction is a segment bounded as in  $r_1<r<r_2$, with $r_1$ the infra-red (IR) boundary and $r_2$ the ultra-violet (UV) boundary. These boundaries have no physical meaning: they are used to introduce regulators in the IR and UV of the dual theory, and should be removed by sending $r_2\rightarrow +\infty$ and $r_1\rightarrow r_o$, where $r_o$ is the end of space of the geometry. The determinant  of the background metric is such that $\sqrt{-g_5}=e^{4A}$, and, evaluated on the background, the vector $N_M$ ortho-normalised to the boundary, in this choice of coordinates, is given by
\beqs
N^M&=&{\rm diag}\,\left(0\,,\,1\right)\,,~~~~~~~N_M\,=\,g_{MN}N^N\,=\,{\rm diag}\,\left(0\,,\,1\right)\,,
\eeqs
so that the induced metric is (see Appendix~\ref{Sec:B1})
\beqs
\tilde{g}_{MN}&\equiv& {\rm diag}\,\big(e^{2A}\eta_{\mu\nu}\,,\,0\big)\,,
\eeqs
while the Gibbons-Hawking term is $K=-4\partial_rA$.

Here and in the rest of the paper, we assume that the background classical solutions of the system in $D=5$ dimensions be characterised only by the metric (namely the function $A(r)$) and by the background scalars $\phi(r)$ and $\chi(r)$, and that they depend only on the radial direction $r$. All other fields are trivial in the background, and Lorentz invariance is ensured by the fact that no background function depends on the four-dimensional coordinates $x^{\mu}$.

\subsubsection{Fixed point solutions}
\label{Sec:fixedsol}

\begin{figure}[t]
\begin{center}
\includegraphics[width=8cm]{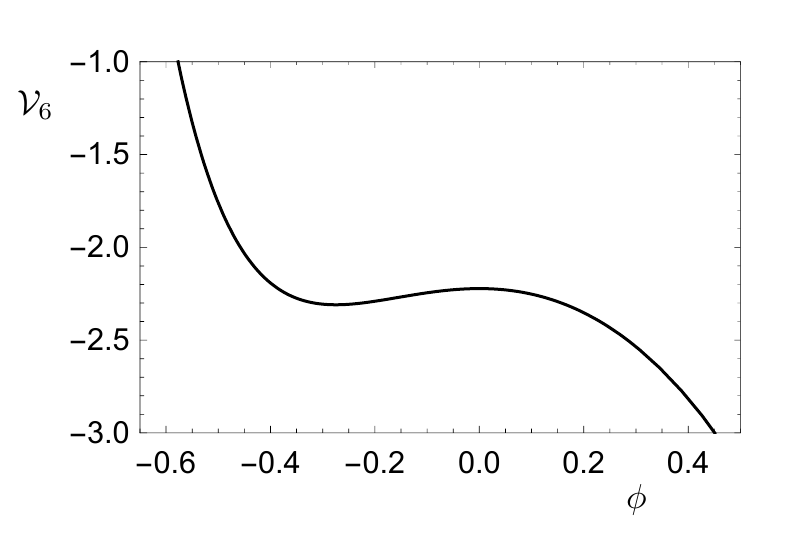}
\caption{The scalar potential $\mathcal{V}_{6}(\phi)$ of the model in $D=6$ dimensions, as a function of the one scalar field $\phi$.}
\label{ScalarPotential(6D)}
\end{center}
\end{figure}

The scalar potential $\mathcal{V}_{6}(\phi)$ in the action of the six-dimensional model is shown in Figure~\ref{ScalarPotential(6D)}.
It  admits the following two critical points:
\begin{align}
\phi_{UV}&=0 \hspace{1.45cm} \rightarrow \-\ \mathcal{V}_{6}(\phi_{UV})=-\frac{20}{9}\,,\\
\phi_{IR}&=-\frac{\log(3)}{4} \-\ \rightarrow \-\ \mathcal{V}_{6}(\phi_{IR})=-\frac{4}{\sqrt{3}}\,,
\end{align}
which correspond to two distinct five-dimensional conformal field theories; the former CFT is supersymmetric whereas the latter is not~\cite{Gursoy:2002tx}. We chose the labels $\phi_{IR,UV}$ to reflect the fact that there exist solutions describing the renormalisation group flow from $\phi_{UV}$ at short distances to $\phi_{IR}$ at long distances, as we will exhibit later. With the same conventions as in \cite{Elander:2013jqa} these two $AdS_{6}$ solutions have curvature radii~\cite{Karndumri:2012vh}:
\begin{align}
R_{UV}^{2}&=-5[\mathcal{V}_{6}(\phi_{UV})]^{-1}=\frac{9}{4}\, ,\\
R_{IR}^{2}&=-5[\mathcal{V}_{6}(\phi_{IR})]^{-1}=\frac{5\sqrt{3}}{4}\, .
\end{align} 
The mass of the scalar in the $AdS_{6}$ bulk may be read off in each case as the coefficient of the term quadratic in $\phi$ in an expansion of the potential $\mathcal{V}_{6}$ around its extrema:
\begin{align}
\mathcal{V}_{6}(\phi_{UV})&\approx-\frac{20}{9}-\frac{8\phi^{2}}{3}+\mathcal{O}(\phi^{3})+\ldots\,,\\
\mathcal{V}_{6}(\phi_{IR})&\approx-\frac{4}{\sqrt{3}}+\frac{8}{\sqrt{3}}(\phi- \phi_{IR})^{2}+\mathcal{O}\big((\phi- \phi_{IR})^{3}\big)+\ldots\,,
\end{align} 
from which we find
\begin{align}
m_{UV}^{2}&=-\frac{8}{3}\,,\\
m_{IR}^{2}&=\frac{8}{\sqrt{3}}\,,
\end{align}
and hence
\begin{align}
m_{UV}^{2}R_{UV}^{2}&=-6\,,\\
m_{IR}^{2}R_{IR}^{2}&=10\,.
\end{align}

The scaling dimension $\Delta$ of the operator that in the $D$-dimensional dual field theory is connected to a scalar supergravity field in the $AdS_{D+1}$ can be computed from the mass $m^{2}R^{2}$ of the latter  via the relation
\be
m^{2}R^{2}=\Delta(\Delta-D)\,,
\ee   
from which we can determine the dimension of the boundary operators dual to $\phi$ in $D=5$ dimensions for each critical point of the scalar potential, to obtain
\be
\Delta_{UV}=3 \-\ , \-\ \Delta_{IR}=\frac{1}{2}\big(5+\sqrt{65}\big)\,,
\ee
where in solving the quadratic equation we kept only the largest root in each case.

\subsubsection{Simple confining solutions}
\label{Sec:cc}

There exist exact analytical solutions of the equations of motion in $D=5$ dimensions with $\phi=\phi_{0}$, where $\phi_{0}$ corresponds to either of the critical point solutions of the scalar potential in $D=6$ dimensions. Defining $v\equiv\mathcal{V}_{6}(\phi_{0})$, these solutions are given by~\cite{Elander:2013jqa}
\begin{align}
\phi&=\phi_{0},\\
\chi&=\chi_{0}+\frac{1}{15}\log(2)-\frac{1}{5}\log\bigg[\cosh\bigg(\frac{{\sqrt{-5v}}}{2}\rho \bigg) \bigg]
+\frac{1}{3}\log\bigg[\sinh\bigg(\frac{{\sqrt{-5v}}}{2}\rho \bigg) \bigg],\\
A&=A_{0}+\frac{4}{15}\log(2)+\frac{4}{15}\log\bigg[\sinh\bigg(\sqrt{-5v}\rho \bigg) \bigg]
+\frac{1}{15}\log\bigg[\tanh\bigg(\frac{{\sqrt{-5v}}}{2}\rho \bigg) \bigg],
\end{align}
where we introduced the radial coordinate $\rho$ defined by $\dd \rho = e^{-\chi} \dd r$, $\chi_0$ and $A_0$ are two integration constants, and we fixed another integration constant so that the space ends at $\r = 0$.

\subsubsection{Interpolating solutions}
\label{Sec:flowssol}

We are mostly interested in a class of solutions for $\phi$, $\chi$ and $A$ that smoothly interpolates between the two confining solutions in Section~\ref{Sec:cc}, and are known numerically. Following~\cite{Elander:2013jqa}, these interpolating solutions form a one-parameter family, characterised by the choice of a parameter $\tilde{\phi}$ that determines the scale at which the flow between the two distinct CFTs transitions from one to the other.

To obtain the interpolating solutions, the classical equations of motion derived from the five-dimensional action $S_{5}$ may  be rewritten  as follows:\footnote{Note that Eq.~\eqref{eq:Hamilder} is not independent, but can be obtained by differentiating the Hamiltonian constraint Eq.~\eqref{eq:Hamil} with respect to $\rho$, and substituting the equations of motions for the scalars $\phi$ and $\chi$.}
\begin{align}
\partial^{2}_{\rho}\phi+(4\partial_{\rho}A-\partial_{\rho}\chi)\partial_{\rho}\phi&=\frac{1}{2}\frac{\partial\mathcal{V}_{6}}{\partial\phi}\, ,\\
\partial^{2}_{\rho}\chi+(4\partial_{\rho}A-\partial_{\rho}\chi)\partial_{\rho}\chi&=-\frac{\mathcal{V}_{6}}{3}\, ,\\
\label{eq:Hamilder}
3\partial^{2}_{\rho}A +6(\partial_{\rho}A)^{2}+2(\partial_{\rho}\phi)^{2}+6(\partial_{\rho}\chi)^{2}-3\partial_{\rho}A\partial_{\rho}\chi
&=-2\mathcal{V}_{6}\, ,\\
\label{eq:Hamil}
3(\partial_{\rho}A)^{2}-(\partial_{\rho}\phi)^{2}-3(\partial_{\rho}\chi)^{2}&=-\mathcal{V}_{6}\, .
\end{align}  

In order to solve the equations numerically, we set up the boundary conditions by making use of the expansion for $\phi$, $\chi$ and $A$ about the end of space at $\r=0$. The one-parameter family of interest generalises the form of the simple confining solutions in such a way that $\phi$ behaves regularly near the $\rho=0$ region, and reads~\cite{Elander:2013jqa}
\begin{align}
\phi &= \big(\tilde{\phi}-\frac{1}{4}\log(3)\big) 
-\frac{e^{-6\tilde{\phi}}}{4\sqrt{3}} 
\big(3 - 4e^{4\tilde{\phi}} + e^{8\tilde{\phi}}\big)\rho^{2}\notag \\
&+ \frac{e^{-12\tilde{\phi}}}{36}\big(-12 + 28e^{4\tilde{\phi}} - 
17e^{8\tilde{\phi}} + e^{16\tilde{\phi}}\big)\rho^{4}+\mathcal{O}(\rho^{6})\, ,\\
\chi&=\chi_{0} + \frac{1}{60}\big(20\log(\rho) + 4\log(2) + 5\log(25/3) \big) \notag\\
&-\frac{e^{-2\tilde{\phi}}}{9\sqrt{3}}\big(\sinh(4\tilde{\phi}) + 2\big)\rho^{2} + 
\frac{5e^{-4\tilde{\phi}}}{162}\big(\sinh(4\tilde{\phi}) + 2\big)^{2}\rho^{4}+\mathcal{O}(\rho^{6})\, ,\\
A&=A_{0} + \frac{1}{60}\big(20\log(\rho) + 32\log(2) + 5\log(25/3) \big)
+\frac{7e^{-2\tilde{\phi}}}{18\sqrt{3}}\big(\sinh(4\tilde{\phi}) + 2\big)\rho^{2} \\
&+ \frac{e^{-4\tilde{\phi}}}{324}
\big(108\cosh(4\tilde{\phi})-2\big(20\cosh(8\tilde{\phi})+52\sinh(4\tilde{\phi})+59\big) +27\sinh(8\tilde{\phi}) \big)\rho^{4}+\mathcal{O}(\rho^{6})\, . \notag
\end{align}    
By imposing boundary conditions on $\phi$, $\chi$ and $A$ (at small $\r$) dictated by these IR expansions,
and solving the background equations, we obtain the desired family of numerical solutions. We constrain the parameter $\tilde{\phi}$ to take values $0\leq\tilde{\phi}\leq\frac{1}{4}\log(3)$. Following~\cite{Elander:2013jqa}, in our analysis we adopt the convenient redefinition:
\be
\tilde{\phi}=\frac{1}{8}\log(3)\Big[1-\tanh\Big(\frac{s_{\ast}}{2} \Big) \Big]\, ,
\ee
so that the limits $s_{\ast}\rightarrow +\infty$ and $s_{\ast}\rightarrow -\infty$ correspond to the limits $\tilde{\phi}\rightarrow 0$ and $\tilde{\phi}\rightarrow \frac{1}{4}\log(3)$, respectively, thus reproducing the simple confining solutions of the previous section. Figure~3 in~\cite{Elander:2013jqa} illustrates a sample of solutions built in this way. We verified explicitly that the six-dimensional backgrounds we use are regular.

\section{The mass spectrum of glueballs}
\label{Sec:Mass}

In this section we present the main results of our numerical analysis. In Section~\ref{Sec:fluctuations} we provide all the equations and boundary conditions obeyed by the physical,  gauge-invariant combinations of fluctuations of the backgrounds of interest. The general expressions for all the equations, and their derivations in the case of $p$-forms, can be found in Appendix~\ref{Sec:B}. In Section~\ref{Sec:fixed} we tabulate the glueball masses computed by fluctuating the gravity backgrounds in which $\phi$ assumes the constant value characterising each critical point of the system in $D=6$ dimensions,  and described in Section~\ref{Sec:fixedsol}. In Section~\ref{Sec:flows} we provide plots of the mass spectra obtained by  numerically solving the fluctuation equations and boundary conditions derived from background solutions which interpolate between the two critical points, in terms of the transition scale parameter $s_{*}$ introduced in Section~\ref{Sec:flowssol}.

\subsection{Equations for the fluctuations}
\label{Sec:fluctuations}

The model defined by the complete five-dimensional action given in Eq.~\eqref{Eq:S5ast} has a number of different gauge invariances, in addition to diffeomorphisms: there is the $U(1)$ associated with the gravi-photon $V_M$, the $SU(2)$ associated with the vectors $A^i_M$ and pseudo-scalars $\pi^i$, as well as the gauge invariance of the two-form $B_{MN}$ and the vector $A_M$, and the $U(1)$ of the vector $B_{6N}$ and pseudo-scalar $A_6$. As explained in Appendix~\ref{Sec:B}, these gauge invariances can be treated separately, due to the fact that all the pseudo-scalars and the $p$-forms vanish on the background solutions, and that the computation of spectra only requires retaining in the action terms up to second order in the fluctuations.

In presenting the equations for the gauge-invariant physical fluctuations to be solved numerically, we use the rescaled holographic coordinate $\rho$ defined earlier on by $\partial_{r}=e^{-\chi}\partial_{\rho}$, and find it convenient to introduce the physical mass $M^2=-q^2$. The three linearised bulk equations for the gauge-invariant scalar fluctuations
$\mathfrak{a}^{a}=\mathfrak{a}^{a}(M,\r)$ are~\cite{Elander:2013jqa}:   
\be
\label{eq:scalarflucs1}
0=\left[\frac{}{}e^{\chi}\mathcal{D}_{\rho}(e^{-\chi}\mathcal{D}_{\rho}) + (4\partial_{\rho}A)\mathcal{D}_{\rho}
+e^{2\chi-2A}M^{2}\right]\mathfrak{a}^{a} - e^{2\chi} \mathcal X^{a}_{\ c}
\mathfrak{a}^{c}\,,
\ee
where
\begin{align}
\label{eq:scalarflucs2}
\mathcal X^{a}_{\ c}=
&-e^{-2\chi}\mathcal{R}^{a}_{\ bcd}
\partial_{\rho}\bar \Phi^{b}
\partial_{\rho}\bar \Phi^{d} 
+ \mathcal{D}_{c}\bigg(G^{ab}\frac{\partial \mathcal{V}}{\partial \bar \Phi^{b}}\bigg)\,+\notag\\
&+\frac{4}{3\partial{\rho} A}
\bigg[\partial_{\rho}\bar \Phi^{a}\frac{\partial \mathcal{V}}{\partial \bar \Phi^{c}}+G^{ab}\frac{\partial \mathcal{V}}{\partial \bar \Phi^{b}}\partial_{\rho}\bar \Phi^{d}G_{dc}\bigg]
+\frac{16\mathcal{V}}{9(\partial_{\rho}A)^{2}}\partial_{\rho}\bar \Phi^{a}\partial_{\rho}\bar \Phi^{b}G_{bc}\,.
\end{align}
The boundary conditions read
{\small
\be
\label{eq:scalarflucs3}
\left.\frac{}{}e^{-2\chi}\partial_{\rho}\bar \Phi^{c}
\partial_{\rho}\bar \Phi^{d}G_{db}\mathcal{D}_{\rho}\mathfrak{a}^{b}\right|_{\rho_{i}}
=-\left.\left[\frac{}{}\frac{3\partial_{\rho}A}{2}e^{-2A}M^{2}\delta^{c}_{\ b}
-\partial_{\rho}\bar \Phi^{c}\bigg(\frac{4\mathcal{V}}{3\partial_{\rho}A}\partial_{\rho}\bar \Phi^{d}G_{db} + \frac{\partial\mathcal{V}}{\partial\bar \Phi^{b}} \bigg) \right]\mathfrak{a}^{b}\right|_{\rho_{i}}\, .
\ee
}
The notation in Eq.~\eqref{eq:scalarflucs1}, Eq.~\eqref{eq:scalarflucs2}, and Eq.~\eqref{eq:scalarflucs3}, as well as the origin of the gauge invariant scalars $\mathfrak{a}^a$, is discussed in Appendix~\ref{Sec:B1}. Here we only remind the reader that these fields result from the mixing of fluctuations of the sigma-model scalars  $\Phi^{a} = \{\phi, \chi, \pi^{i}\}$ with the scalar components of the fluctuations of the metric.

The transverse part of the gravi-photon $V_{\mu}=V_{\mu}(M,\r)$ obeys the bulk equations
\be
0 = P^{\mu\nu}\left[\frac{}{}e^{-\chi}\partial_{\rho}\left(\frac{}{}e^{2A}He^{-\chi}\partial_{\rho}V_{\nu}\right)+M^{2}H V_{\nu}\right]\, ,
\ee
where $H=\frac{e^{8\chi}}{4}$, $P^{\mu\nu}$ is the projector defined in Eq.~(\ref{Eq:projector}), 
and the boundary conditions are 
\be
P^{\mu\nu}\,\partial_{\rho}V_{\nu}\big|_{\rho_{i}}=0\, .
\ee
The transverse polarisations of the $SU(2)$ vectors $A^{i}_{\mu}=A^{i}_{\mu}(M,\r)$ obey the same equations and boundary conditions as for the gravi-photon, but with the replacement $H=e^{2\chi-2\phi}$.

The transverse, traceless part of the tensor fluctuations $\mathfrak{e}^{\mu}_{\ \nu}=\mathfrak{e}^{\mu}_{\ \nu}(M,\r)$ obey the bulk equations
\be
0=\left[\frac{}{}\partial^{2}_{\rho}+(4\partial_{\rho}A-\partial_{\rho}\chi)\partial_{\rho}+e^{2\chi-2A}M^{2}\right]\mathfrak{e}^{\mu}_{\ \nu}\, ,
\ee
and the boundary conditions
\be
\partial_{\rho}\mathfrak{e}^{\mu}_{\ \nu}\big\rvert_{\rho_{i}}=0\, .
\ee

We now consider the $U(1)$ gauge fields and the components of the 2-form $B_{MN}$. We start with the sub-system consisting of $B_{6\mu}$, $B_{65}$ and $A_{6}$. For the transverse polarisation of the vector $B_{6\mu}=B_{6\mu}(M,\r)$, the bulk equations are
\be
0=
\left[\frac{}{}-M^{2}-e^{3\chi-4\phi}
\partial_{\rho}(e^{2A-5\chi+4\phi}\partial_{\rho})+m^{2}e^{2A-2\chi-6\phi}\right]P^{\mu\nu}B_{6\nu}
\ee
subject to the boundary conditions
\be
P^{\mu\nu} \, \partial_{\rho} B_{6\nu}\big|_{\rho_{i}}=0\,,
\ee
having set the constants  $D_{i}=0=C_{i}$ in Eq.~(\ref{Eq:bcU15D}).
To decouple the scalar fluctuations $B_{65}$ and $A_{6}$ we rewrite the equations in terms of a new gauge-invariant field $X=X(M,\r)$ defined by
\be
B_{65}\equiv e^{-4A+6\chi+2\phi}X-\frac{1}{m}e^{-\chi}\partial_{\r}A_{6}\,,
\ee
as explained in Appendix~\ref{Sec:B2}. We then obtain the following bulk equation:
\be
0=\partial^{2}_{\r}X+\left(\frac{}{}-2\partial_{\r}A+2\partial_{\r}\phi+5\partial_{\r}\chi\right)\partial_{\r}X
-\left(\frac{}{}-M^{2}e^{-2A+2\chi}+m^{2}e^{-6\phi}\right)X\, ,
\ee
subject to the boundary condition
\be
X\big|_{\rho_{i}}=0\,,
\ee
where we again set $C_{i}=0$ in Eq.~(\ref{Eq:bcXU15D}), reducing the boundary conditions to Dirichlet.

Finally, we consider the sub-system consisting of $A_{\mu}$, $A_{5}$, $B_{\mu\nu}$ and $B_{5\mu}$, following the procedure outlined in Appendix~\ref{Sec:B3}; the six degrees of freedom in this sub-system can be thought of as describing a massive 2-form $B_{\mu\nu}$ and a massive vector $X_{\mu}=X_{\mu}(M,\r)$ defined by
\be
B_{5\mu}\equiv e^{-2A-2\chi+2\phi}X_{\mu}-\frac{1}{m}e^{-\chi}\partial_{\r}A_{\mu}\,.
\ee 
The bulk equations for the transverse polarisations of $B_{\mu\nu}=B_{\mu\nu}(M,\r)$ and $X_{\mu}$ are
{\small
\begin{align}
0&=P^{\mu\rho}P^{\nu\sigma}\big[M^{2}e^{-2A}+e^{-5\chi-4\phi}\partial_{\rho}\big(e^{3\chi+4\phi}\partial_{\rho} \big) 
-m^{2}e^{-2\chi-6\phi} \big]B_{\rho\sigma}
\, , \\
0&=P^{\mu\nu}\big[e^{-\chi}\partial_{\rho}
\big(e^{-\chi}\partial_{\rho}X_{\nu} \big)
-(2\partial_{\rho}\chi-2\partial_{\rho}\phi)e^{-2\chi}\partial_{\rho}X_{\nu}
+(e^{-2A}M^{2}-m^{2}e^{-2\chi-6\phi} )X_{\nu} \big]\,,
\end{align}
}
and the corresponding boundary conditions are 
\beqs
	0 &=& P^{\mu\tau} P^{\nu\sigma} \partial_\r B_{\tau\sigma} \big |_{\r_i} \, ,\\
	 0&=& P^{\mu\nu} X_{\nu} \big |_{\r_i} \,,
\eeqs
where we set the  parameters $D_{i}=0=E_{i}$ in Eqs.~(\ref{Eq:bc2form5d}) and~(\ref{Eq:bc1form5D}), and hence  reduced the boundary conditions to Neumann and Dirichlet for the 2-form and 1-form, respectively.

\begin{table}
\begin{center}
\begin{small}
\begin{tabular}{|ccccccccc|}
\hline\hline
Spin-0 & Spin-1 & Spin-2 & Spin-0 & Spin-1 & Spin-0 & Spin-1 & Spin-1 & Spin-1 \cr
$\mathfrak{a}^a$ &  $V_{\mu}$ & $\mathfrak{e}^{\mu}_{\ \nu}$ & $\pi^i$ & $A^{i}_{\mu}$ & $X$&$B_{6\nu}$ & $X_{\mu}$ &$B_{\mu\nu}$\cr
\hline
 0.54 & 1.23 & 1.00 & 1.00 & 0.73 & 0.60 & 0.40 & 1.02 & 0.66 \\
 0.62 & 1.91 & 1.65 & 1.65 & 1.38 & 1.35 & 1.07 & 1.66 & 1.34 \\
 1.15 & 2.55 & 2.28 & 2.28 & 2.00 & 2.00 & 1.72 & 2.29 & 1.98 \\
 1.53 & 3.18 & 2.90 & 2.90 & 2.63 & 2.64 & 2.35 & 2.91 & 2.60 \\
 1.77 & 3.81 & 3.53 & 3.53 & 3.25 & 3.27 & 2.97 & 3.53 & 3.22 \\
 2.20 &  \text{} & \text{} & \text{} & 3.87 & 3.89 & 3.60 &  \text{}  &3.84   \\
 2.39 & \text{} & \text{} & \text{} & \text{} & \text{} & \text{} & \text{} & \text{} \\
 2.84 & \text{} & \text{} & \text{} & \text{} & \text{} & \text{} & \text{} & \text{} \\
 3.01 & \text{} & \text{} & \text{} & \text{} & \text{} & \text{} & \text{} & \text{} \\
 3.48 & \text{} & \text{} & \text{} & \text{} & \text{} & \text{} & \text{} & \text{} \\
 3.64 & \text{} & \text{} & \text{} & \text{} & \text{} & \text{} & \text{} & \text{} \\
\hline\hline
\end{tabular}
\end{small}
\end{center}
\caption{Masses $M$ of the first few excitations in all 10 towers of states, normalised to the mass of the lightest tensor mass, computed on backgrounds with $\phi=\phi_{UV}=0$. The numerical calculations are performed by setting the IR cutoff to $\r_1=0.001$, and the UV cutoff to $\r_2=8$. The numerical solutions are obtained by the midpoint determinant method, computed at the intermediate $\r_{\ast}=4$.}
\label{Fig:UVspectrum}
\end{table}

\begin{table}
\begin{center}
\begin{small}
\begin{tabular}{|ccccccccc|}
\hline\hline
Spin-0 & Spin-1 & Spin-2 & Spin-0 & Spin-1 & Spin-0 & Spin-1 & Spin-1 & Spin-1 \cr
$\mathfrak{a}^a$ &  $V_{\mu}$ & $\mathfrak{e}^{\mu}_{\ \nu}$ & $\pi^i$ & $A^{i}_{\mu}$ & $X$&$B_{6\nu}$ & $X_{\mu}$ &$B_{\mu\nu}$\cr
\hline
 0.62 & 1.23 & 1.00 & 1.00 & 0.73 & 1.08 & 0.82 & 1.48 & 1.10 \\
 1.44 & 1.90 & 1.65 & 1.65 & 1.37 & 1.82 & 1.54 & 2.13 & 1.80 \\
 1.53 & 2.55 & 2.28 & 2.28 & 2.00 & 2.49 & 2.19 & 2.77 & 2.45 \\
 2.11 & 3.18 & 2.90 & 2.90 & 2.62 & 3.13 & 2.83 & 3.40 & 3.08 \\
 2.20 & 3.81 & 3.53 & 3.53 & 3.25 & 3.76 & 3.46 & \text{}& 3.71  \\
 2.76  & \text{} & \text{} & \text{} & 3.87 & \text{} & \text{} & \text{} & \text{}\\
 2.84 & \text{} & \text{} & \text{} & \text{} & \text{} & \text{} & \text{} & \text{} \\
 3.39 & \text{} & \text{} & \text{} & \text{} & \text{} & \text{} & \text{} & \text{} \\
 3.48 & \text{} & \text{} & \text{} & \text{} & \text{} & \text{} & \text{} & \text{} \\
\hline\hline
\end{tabular}
\end{small}
\end{center}
\caption{Same as Table~\ref{Fig:UVspectrum}, but with $\phi=\phi_{IR}=-\frac{\log 3}{4}$.}
\label{Fig:IRspectrum}
\end{table}

\begin{figure}[t]
\begin{center}
\includegraphics[width=14cm]{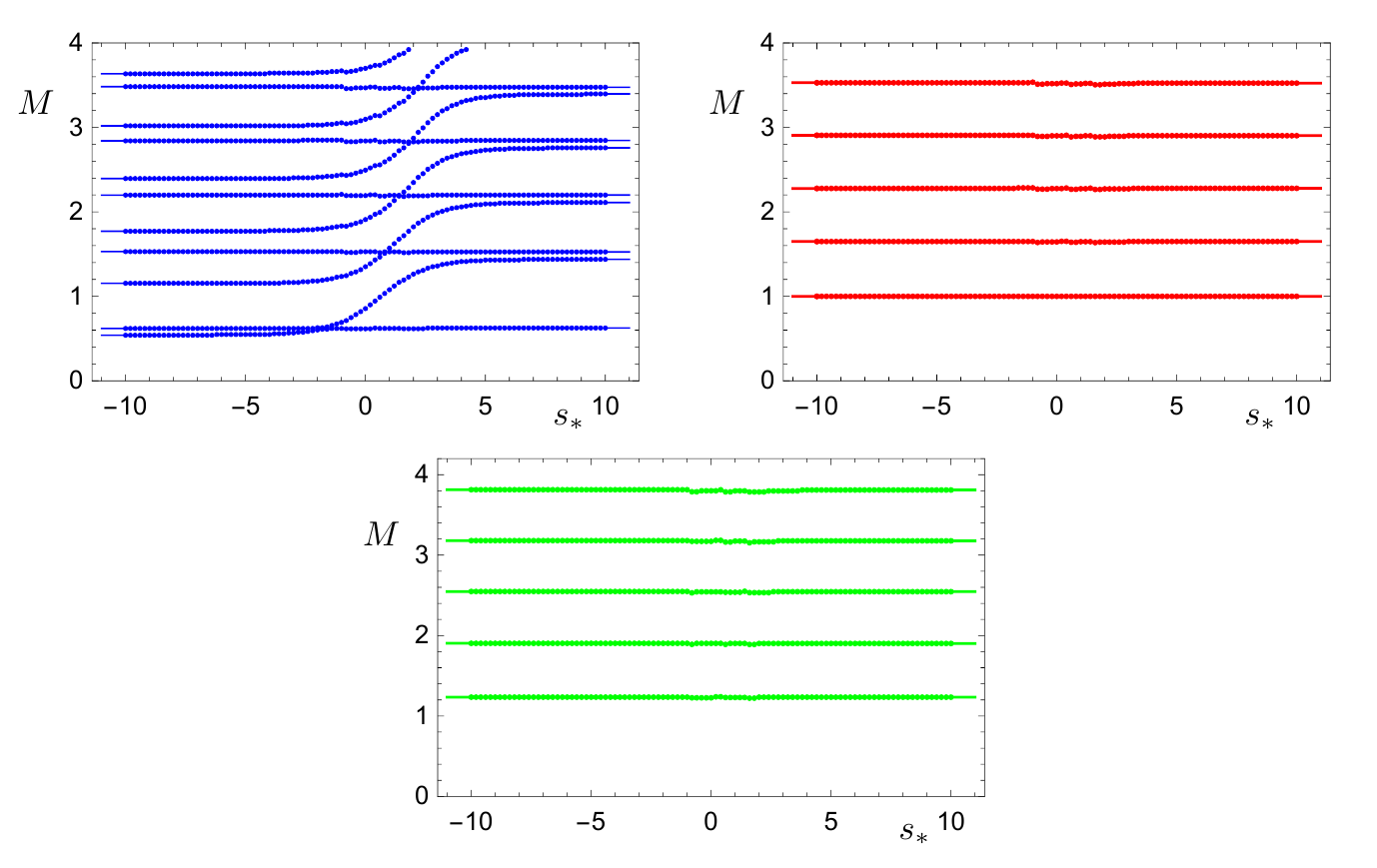}
\end{center}
\caption{The spectrum of masses $M$ as a function of the scale $s_{\ast}$, normalised in units of the mass of the lightest tensor. In all plots, the left and right margins correspond to the spectra computed from the analytical solutions obtained by deforming the six-dimensional critical points. From top to bottom, left to right, the spectra of fluctuations of the two scalars $\chi$ and $\phi$ (blue), tensors  $\mathfrak{e}^{\mu}_{\ \nu}$ (red) and the gravi-photon $V_{\mu}$  (green). The numerical calculations are performed by setting the IR cutoff to $\r_1=0.001$, and the UV cutoff to $\r_2=8$. In the midpoint determinant method we set $\r_{\ast}=2$.
\label{Fig:spectra1}}
\end{figure}

\begin{figure}[t]
\begin{center}
\includegraphics[width=14cm]{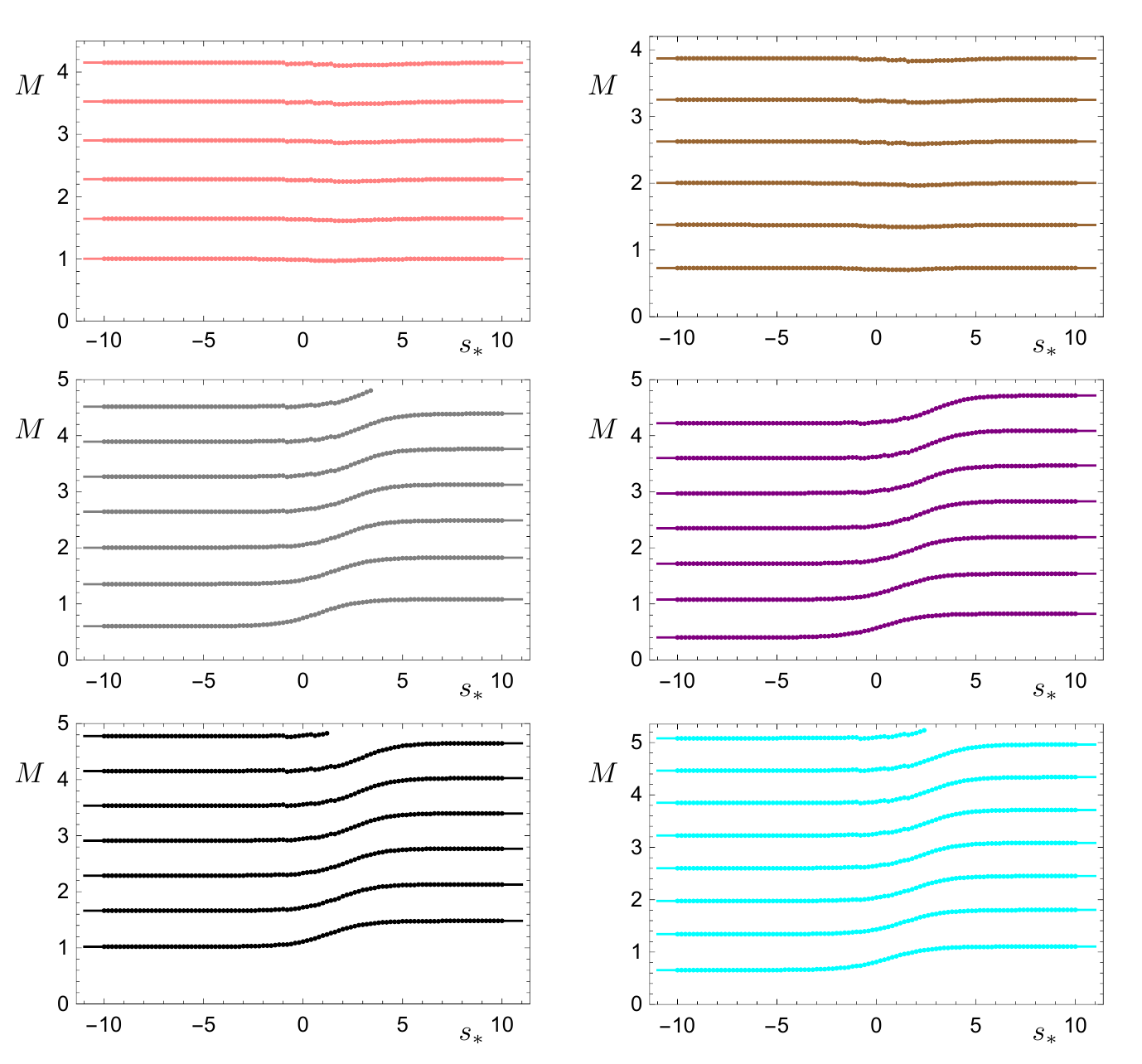}
\end{center}
\caption{The spectrum of masses $M$ as a function of the scale $s_{\ast}$, normalised in units of the mass of the lightest tensor. In all plots, the left and right margins correspond to the spectra computed from the analytical solutions obtained by deforming the six-dimensional critical points. From top to bottom, left to right, the spectra of fluctuations of $SU(2)$ adjoint pseudo-scalars $\pi^{i}$ (pink), the $SU(2)$ adjoint vectors $A^{i}_{\mu}$ (brown), $U(1)$ (pseudo-)scalar $X$ obtained as gauge-invariant combination of  $A_{6}$ and $B_{65}$ (grey), the $U(1)$ transverse vector $B_{6\mu}$ (purple), the $U(1)$ transverse vector $X_{\mu}$ (black) and the massive $U(1)$ 2-form $B_{\mu\nu}$ (cyan). The numerical calculations are performed by setting the IR cutoff to $\r_1=0.001$, and the UV cutoff to $\r_2=8$, and in the midpoint determinant method we set $\r_{\ast}=2$.
\label{Fig:spectra2}}
\end{figure}

\subsection{Mass spectra for simple confining solutions}
\label{Sec:fixed}

We summarise in Tables~\ref{Fig:UVspectrum} and~\ref{Fig:IRspectrum} our numerical results for the spectra of modes computed, respectively, for the analytical background solutions with $\phi=\phi_{UV}=0$ and $\phi=\phi_{IR}=-\frac{\log 3}{4}$.
We restrict to the first few such states. The procedure adopted in the numerics employs the mid-determinant method: for each value of the trial mass squared $M^2$, we impose independently the IR and UV boundary conditions on the solutions to the linearised bulk equations, and evolve them to a mid-point $\r_{\ast}$ in the radial direction $\r$. We construct  the matrix of the 
resulting fluctuations and their derivatives, evaluated at $\r_{\ast}$, including both the solutions evolved from the IR and from the UV, and compute the determinant. By  varying the trial value of $M^2$, we look for the zeros of this determinant.

We chose values of $\r_1$ and $\r_2$ in such a way as to ensure that the results for the spectra are independent of the position of the regulators. We report as final results the numerical values of $M$ obtained for the same choices of cut-offs $\r_i$ adopted
in~\cite{Elander:2013jqa}, and we verified that (for states for which the comparison is possible) our results agree with those in~\cite{Elander:2013jqa}. We also considered negative values of $M^2$: the absence of tachyonic modes supports the perturbative stability of the solutions, also in the presence of the circle compactification.

In order to facilitate comparison between spectra of states with different spin,  and with results from other papers in the literature, in this paper we normalised the whole spectrum to the mass of the lightest particle of spin-2 (tensor mode).

\subsection{Mass spectra for interpolating solutions}
\label{Sec:flows}

The numerical calculations of the spectra in the more general case in which the function $\phi$ is allowed to evolve between its two critical values follows the same procedure as for the case in which $\phi$ is constant. The only difference is that in this case the background solutions are known only numerically. We generated a large set of numerical solutions of the background equations, each of which is characterised by a different value of $s_{\ast}$ as defined in Section~\ref{Sec:flowssol}, and applied to them the process for calculating the fluctuations. We show the results in Figures~\ref{Fig:spectra1} and~\ref{Fig:spectra2}, which are obtained by making use of the same parameters in the numerical calculations as  in Section~\ref{Sec:fixed}.

Notice that at the furthest left region of each of the individual panels in Figures~\ref{Fig:spectra1} and~\ref{Fig:spectra2} the numerical solutions are compared to the case $s_{\ast}\rightarrow -\infty$ computed in Section~\ref{Sec:fixed}, while in the furthest right region they are compared to the case $s_{\ast}\rightarrow +\infty$. In this way we checked that indeed the numerical calculations converge to the correct asymptotic values. For a large value $\rho_2=12$ of the UV cutoff, the numerical results do not show any appreciable difference with the $\r_2=8$ case.

\section{Discussion}
\label{Sec:Discussion}

\begin{figure}[t]
\begin{center}
\includegraphics[width=13.7cm]{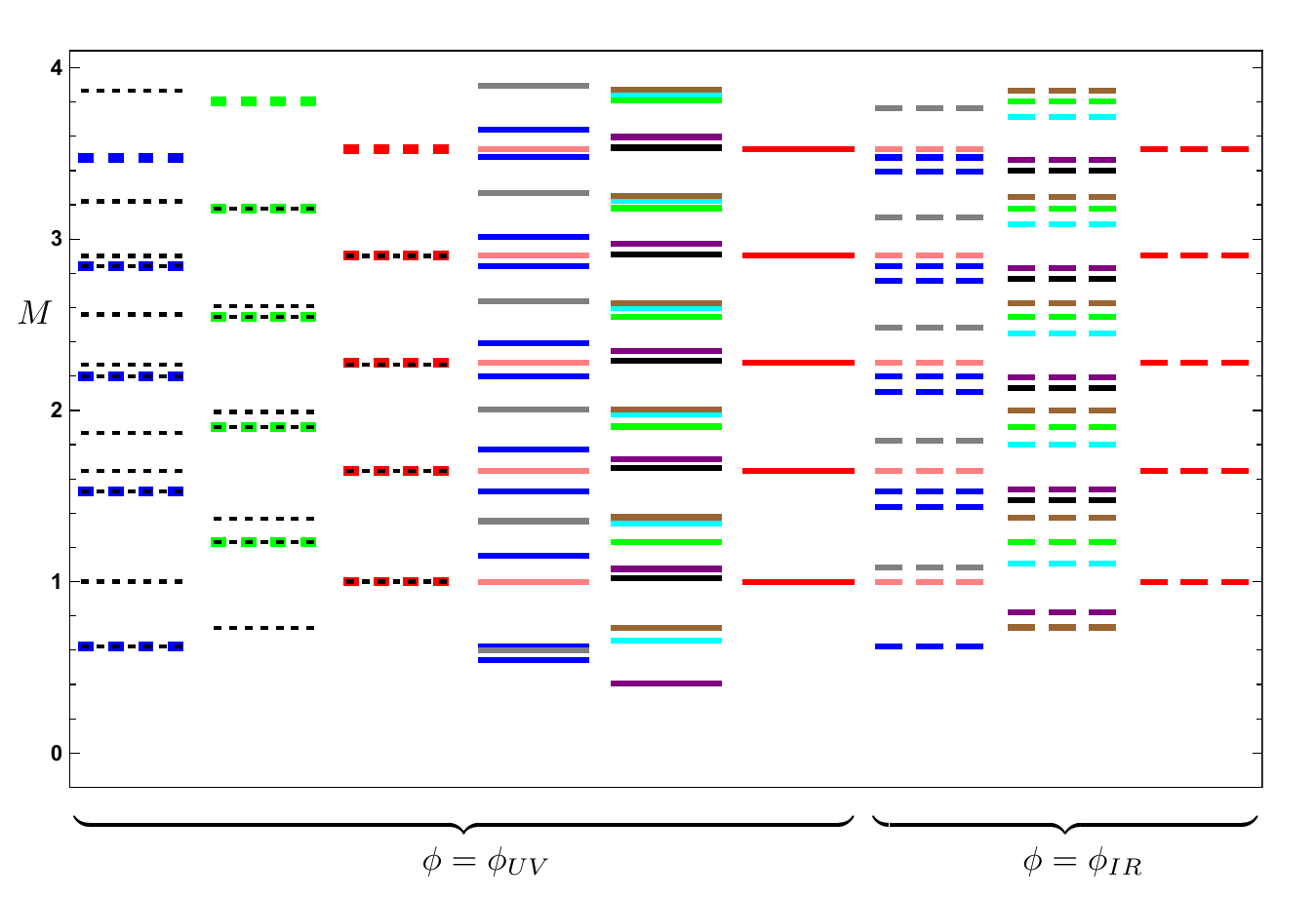}
\end{center}
\caption{Masses of two scalars associated with $\chi$ and $\phi$ (blue), scalar adjoint of $SU(2)$ $\pi^i$ (pink), $U(1)$ scalar $X$ obtained as gauge-invariant combination of  $A_{6}$ and $B_{65}$ (grey), gravi-photon $V_{\mu}$  (green), vectorial adjoint of $SU(2)$ $A_{\mu}^i$ (brown), $U(1)$ vectorial $B_{6\mu}$ (purple), $U(1)$ vector $X_{\mu}$ gauge-invariant combination of $B_{5\mu}$ and $A_{\mu}$ (black), $U(1)$ 2-form $B_{\mu\nu}$ (cyan), and tensorial (red) fluctuations, for the model obtained from the circle compactification of the $D=6$ supergravity theory, as computed in~\cite{Wen:2004qh} (short dashed), by the authors of~\cite{Kuperstein:2004yf} (dotted, black) and by us (continuous and long dashed). In our calculation, we retain one extra scalar mode compared to~\cite{Wen:2004qh}, corresponding to fluctuations of $\phi$, the spectrum  of which depends on the value of $\phi_0$. The three towers in the middle (continuous) of the figure are the spectrum obtained for $\phi=\phi_{UV}=0$ (see also Table~\ref{Fig:UVspectrum}), while the three rightmost towers (long-dashed) are the spectrum for $\phi=\phi_{IR}=-\frac{1}{4}\log(3)$ (see also Table~\ref{Fig:IRspectrum}). In the calculations, we fixed $\r_1=0.001$ and $\r_2=8$. Notice that the spectrum from~\cite{Kuperstein:2004yf} contains six towers: three of them agree with~\cite{Wen:2004qh} as well as us, one agrees with the vector $SU(2)$ fields  from our calculation, one scalar is degenerate with the tensor, and agrees with our $SU(2)$ adjoint scalars $\pi^i$, up to small numerical discrepancies.}
\label{Fig:spectraF(4)CONF}
\end{figure}

We start this discussion session with a general observation pertaining to the nature and properties of the 10 towers of states we analysed in the one-parameter class of models of this paper. In~\cite{Elander:2013jqa} it was observed that the fluctuations of the scalar that we call $\chi$ have a {\it universal} character, in the sense that they appear in a large class of supergravity backgrounds, and their masses are not sensitive to specific details. Evidence collected in this paper extends this observation to the fluctuations of the graviton and of the gravi-photon that, as shown in Figure~\ref{Fig:spectra1}, are unaffected by the background choice within the one-parameter class of classical background solutions we studied. All these  modes  descend from the reduction on a circle of the six-dimensional graviton.

We compare our results to those in the literature for related backgrounds. The earliest analysis we found in the literature of the mass spectrum within this class of models is restricted to only the three universal towers discussed above, for which the results are summarised in Table~1 of~\cite{Wen:2004qh}. The authors considered only the background for which $\phi=0$ (or equivalently, in our notation, $s_{\ast}\rightarrow -\infty$), hence allowing only the deformation of the dual CFT that is described within the gravity theory by the compactification of the direction $\eta$, but without flowing between the fixed points. In Figure~\ref{Fig:spectraF(4)CONF}, the blue, green, and red dashed lines show the numerical results from~\cite{Wen:2004qh} (three leftmost columns), compared with ours in the same background (middle three columns) as well as in the background in which $\phi$ assumes the value of the IR fixed point of the dual five-dimensional gauge theory (three rightmost columns). The three sets are in agreement, within the numerical resolution, for all three towers of universal states. Compared to~\cite{Wen:2004qh}, in this work we show explicitly that these masses are independent of $s_{\ast}$.

The backgrounds with $s_{\ast}\rightarrow -\infty$ have also been analysed in~\cite{Kuperstein:2004yf}, that reports on a larger set of modes that includes six towers of states, two of which (one of the spin-0 and the spin-2) happen to be degenerate in mass.
These six towers are reported in our Figure~\ref{Fig:spectraF(4)CONF}, on the three left-most columns, as dotted black lines.
The three universal states agree both with the calculation in this paper and that in~\cite{Wen:2004qh}. Two of the towers in~\cite{Kuperstein:2004yf} are obtained  by fluctuating a RR 1-form, which yields one tower of pseudo-scalar and one of vector modes. The resulting towers agree within the numerical resolution with our results for the $SU(2)$ triplets, both in the case of the scalar and of the vector (barring the three-fold degeneracy,) but not with any of the states in the system formed by the
(massive) $U(1)$ vector and the 2-form. Our analysis extends the results to the whole one-parameter family of solutions: these two towers of masses once again show no appreciable dependence of the background chosen, as shown by the two  top panels of Figure~\ref{Fig:spectra2}.

We performed also the calculation of the spectrum of the system given by the six-dimensional massive 2-form and the $U(1)$ six-dimensional vector, for which our results for the four towers of modes are shown in the four bottom panels of Figure~\ref{Fig:spectra2}, a calculation which has not been previously attempted. These states show a non-universal behavior: the fact that the mass term depends on $\phi$ affects the spectrum in a visible way, with all four towers becoming heavier when the background has non-trivial $\phi$.

Another significant difference with Ref.~\cite{Kuperstein:2004yf} appears in the scalar sector: we do not find the heavy tower of scalar states described there, but rather an additional tower of scalar states that starts at moderately light values. It corresponds mostly to fluctuations of $\phi$,  the mass of which depends appreciably on the value of $s_{\ast}$, as shown by the first panel of Figure~\ref{Fig:spectra1}. We observe that $\phi$ is the only scalar field in the supergravity action in $D=6$ dimensions, and that the ten-dimensional dilaton $\Phi$ is a non-trivial function of $\phi$ and of the warp factors appearing in the lift, which in general depend also on one of the coordinates of the internal space~\cite{Cvetic:1999un}. There is hence no other state in the six-dimensional supergravity that can be matched to the heavy scalar tower in~\cite{Kuperstein:2004yf}.

\begin{figure}[t]
\begin{center}
\includegraphics[width=13.7cm]{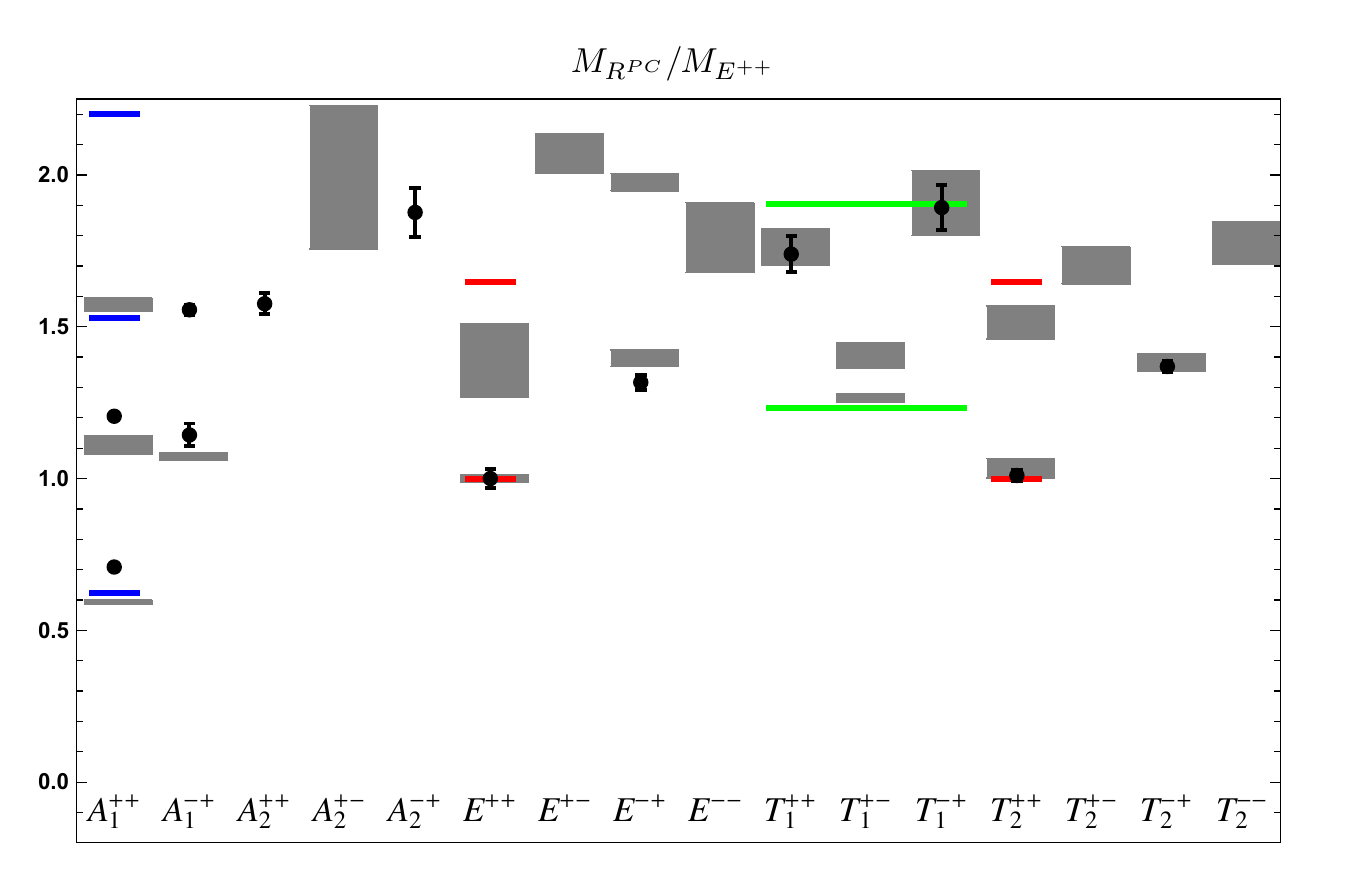}
\end{center}
\caption{Masses of glueballs from lattice results compared to the three towers of {\it universal} results in this paper. Lattice results for $SU(N)$,  extrapolated to $N\rightarrow +\infty$ are taken from Figure~20 in~\cite{Lucini:2010nv}, are labelled left-to-right $A_1^{++},A_1^{+-},\cdots$, the discrete lattice quantum numbers $R^{PC}$, are represented by the shaded grey rectangles,
and are normalised with respect to the $E^{++}$ state as $M_{R^{PC}}/M_{E^{++}}$. Lattice results for $Sp(4)$ are taken from~\cite{Bennett:2017kga}, in which case all states have $C=+$, and are represented by the black dots (with error bars). Notice that we interchanged the $T_1$ and $T_2$ labels, to be consistent with the conventions in~\cite{Lucini:2010nv}. The three universal towers are: the fluctuations of the scalar $\chi$ (blue), the tensors (red), and the gravi-photon (green). We do not commit to a choice of which tower of spin-1 states in the lattice results should be identified with the gravi-photon. We normalise the supergravity masses to the mass of the lightest tensor, so that all three data sets agree on the lightest $E^{++}$ state. }
\label{Fig:plotCompareLRRSp4}
\end{figure}

Because of the fact that the supergravity models we discussed are dual to field theories which resemble Yang-Mills theories in $D=4$ dimensions, at least at large distances,  in particular in reference to the physics of confinement, it is illustrative to compare with lattice calculations. We restrict the comparison to  the results for the three universal towers, for two main reasons. In the first place, Yang-Mills gauge theories in $D=4$ dimensions are entirely characterised by one dynamical scale, and there is no quantity that can be naturally associated  with the parameter $s_{\ast}$,  hence forcing us to exclude from the comparison all the states that show a dependence on $s_{\ast}$ in their masses (see Figures~\ref{Fig:spectra1} and~\ref{Fig:spectra2}). Also, as there is no $SU(2)$ global symmetry in Yang-Mills theories in $D=4$, there is no comparison to make for the $SU(2)$ triplets (both the pseudo-scalar and the vector), and we must exclude these two towers.

The semi-classical calculations performed in the context of gauge-gravity dualities are expected to correspond to the large-$N$ limit of a field theory. For $SU(N)$, we compare to the extrapolation to large $N$ that has been performed in~\cite{Lucini:2010nv}. For $Sp(2N)$ and $SO(N)$ gauge theories, such a systematic study has not been performed yet, and we rely on 
the largest $N$ for which data is available, namely $Sp(4)$ from~\cite{Bennett:2017kga} (which is locally isomorphic to $SO(5)$).
 
An additional difficulty of a technical nature emerges when comparing to lattice data: at finite lattice spacing, the continuum rotation group is broken to a discrete subgroup, which in the case of cubic lattices as in~\cite{Lucini:2010nv} and~\cite{Bennett:2017kga} is the octahedral group. The correspondence between spin $J$ and the five irreducible representations $A_1$, $A_2$, $E$, $T_1$ and $T_2$ is non-trivial,\footnote{There is a discrepancy in the conventions used in~\cite{Lucini:2010nv} and~\cite{Bennett:2017kga}, where the roles of $T_1$ and $T_2$ are interchanged. Here, we follow the conventions and notation of the former.} and we report it in Table~\ref{Fig:J-R}, which we borrow from~\cite{Lucini:2010nv}.

\begin{table}
\begin{center}
\begin{tabular}{|c|ccccc|}
\hline\hline
$~~~~~~~~~~~~~R~~~~$ &$A_1$ & $A_2$ & $E$ &  $T_1$ & $T_2$\cr
$J$ &&  && & \cr
\hline
$0$ & $1$ & $0$ & $0$ & $0$ & $0$ \cr
$1$ & $0$ & $0$ & $0$ & $1$ & $0$ \cr
$2$ & $0$ & $0$ & $1$ & $0$ & $1$ \cr
$3$ & $0$ & $1$ & $0$ & $1$ & $1$ \cr
$4$ & $1$ & $0$ & $1$ & $1$ & $1$ \cr
\hline\hline
\end{tabular}
\end{center}
\caption{Subduced representations $R$ of the octahedral group
in terms of the continuum representations $J$ of the rotational group,
from~\cite{Lucini:2010nv}.}
\label{Fig:J-R}
\end{table}

It is customary also to classify states in terms of the eigenvalues $\pm 1$ of parity $P$ and charge-conjugation $C$, so that each lattice state can be assigned to one of $20$ possible irreducible representations $R^{PC}$, with the caveat that in the case of $Sp(4)$, for which all representations are pseudo-real, $C=+1$ for all states.

In Figure~\ref{Fig:plotCompareLRRSp4} we compare the three towers of universal states identified in this paper with the corresponding lattice states obtained by extrapolating $SU(N)$ to $N\rightarrow +\infty$, taken from Figure~20 in~\cite{Lucini:2010nv}. The grey boxes have  sizes  determined by the statistical error. The systematic errors, particularly in the extrapolation to large $N$, are unknown. Having normalised the states so that the lightest tensor from supergravity agrees with the $E^{++}$ lattice state, the lightest spin-$0$ and spin-$1$ states we computed are just outside of the 1$\sigma$ error bars taken from the lattice.

We also compare to the $Sp(4)$ calculation from~\cite{Bennett:2017kga}, that in Figure~\ref{Fig:plotCompareLRRSp4} is represented by the dots (with statistical errors shown).  The scalar states from the supergravity and lattice results are close to each other, but outside the error bars. Compared to the $SU(N)$ case, the discrepancy has opposite sign, which might be an indication of the fact that the systematics of the large-$N$ extrapolation are not negligible, and of the fact that $Sp(4)$ might still be far from the large $N$ limit. Future measurements of the spectra with larger $Sp(2N)$ groups will help clarify this point. Notice that the spin-1 states cannot be directly compared, suggesting that in the supergravity calculation the $Sp(2N)$ dual might require orbifolding the internal space (along the lines of~\cite{Witten:1998xy}), in a way that would remove part of the spectrum, including this tower of states.

We conclude the comparison with lattice data with an additional comment, mostly driven by the numerical results. It is somewhat intriguing to observe that the scalar $SU(2)$ triplet states in the top-left panel of Figure~\ref{Fig:spectra2} are approximately degenerate with the tensor state, and this is a feature that is not dissimilar to what the lattice data show. Yet, interpreting these states as representative of the pseudo-scalar glueballs would require to include in the comparison also the associated triplet of vectors, which are significantly  lighter than any other spin-1 states on the lattice, suggesting that this observation is probably just due to accidental circumstances.

In order to perform the calculations  presented in this paper, we addressed explicitly some technical subtleties related with gauge invariance  in the presence of $p$-forms. This technical work is of general relevance, as it sets the ground for future work, and we decided to report  upon it in Appendix~\ref{Sec:B}, which also contains extensive discussions. In particular, in the treatment of $p$-forms we show explicitly the boundary-localised terms that are required for holographic renormalisation. As long as we are interested only in the composite states of the theory and their masses, omitting such terms does not alter the results, and this is what we restricted our attention to, in the main body of the paper. Yet, in case one is interested in computing the full 2-point functions (in particular the decay constants), such terms must be included. Furthermore, there are finite ambiguities in the definition of the subtractions that are implicit in the use of the localised terms as counter-terms to remove divergences of the theory in the $r_2\rightarrow +\infty$ limit. There is a subtle connection between these and the possibility of weakly gauging the global symmetries of the dual field theory, which would alter the spectra we computed by reinstating the presence of massless modes which in our analysis are not part of the physical spectrum.

We conclude with another remark. Notice that we did not compute the string tension. It is known that the models considered here provide a description of confinement in terms of a linear potential between static quark sources, and that this can be computed by considering the lift to $D=10$ dimensions and then computing the minimal surface described by open strings with end points localised at the UV boundary, along the usual prescriptions of gauge-gravity duality~\cite{Rey:1998ik,Maldacena:1998im}. However, because the warp factors in the lift involve non-trivially one of the internal angles, in the case of solutions with finite $s_{\ast}$, the resulting system requires solving a non-trivial system of coupled equations~\cite{Elander:2013jqa}. Since the results are of limited interest for our present purposes, we leave this problem for future studies.

\section{Conclusions and Outlook}
\label{Sec:Outlook}

The models we discussed in this paper provide controllable examples of gravity duals of  confining four-dimensional theories that reproduce semi-quantitatively the features of confining Yang-Mills gauge theories. We studied the spectrum of fluctuations of gravity and $p$-forms with $p=0,\,1,\,2$,  by explicitly working out the general  $R_{\xi}$ gauge for all the forms and focusing only on gauge-invariant states, as a function of the parameter $s_{\ast}$ governing the renormalisation group
flow in the higher-dimensional field theory that the current models descend from.

Both the comparison to other supergravity duals, as well as lattice calculations, confirm that the class of background solutions studied in this paper exhibits several qualitative features that make its dual resemble closely the confining dynamics of Yang-Mills theories in $D=4$ dimensions, in spite of the fact that the microscopic theories dual to these supergravity backgrounds are different from that of Yang-Mills. All bound states are characterised by the same scale, including states that carry non-trivial $SU(2)$ and $U(1)$ global quantum numbers. A subset of the particles have masses that do not depend on the details of the background, and in particular on the parameter $s_{\ast}$, suggesting that they are only sensitive to the confinement mechanism, and not the details of the complete model. Conversely, we find explicit evidence of states the masses of which increase when $s_{\ast}$ is non-trivial.

It would be interesting to perform similar calculations in models with different dynamics, that are relevant for light dilaton dynamics or composite-Higgs physics. Among the former, background geometries related to the conifold, such as the baryonic branch of the Klebanov-Strassler system, are of interest. The study of the spectrum of vectors and pseudo-scalar particles within the consistent truncations of~\cite{Cassani:2010na,Bena:2010pr} would provide useful information to better understand the complete symmetry (and supersymmetry) structure of the theory.

In the composite-Higgs context, it would be  interesting to find supergravity backgrounds encompassing one of the patterns of spontaneous symmetry breaking that are employed for model-building purposes. This would substantially differ from the models in which global symmetry and symmetry-breaking are described in terms of a set of extended objects treated in probe approximations, along the lines of what is done in the $D3-D7$~\cite{Karch:2002sh} or $D4-D8$~\cite{Sakai:2004cn} systems, which is more closely related to the treatment of mesons in gauge theories with quenched matter fields.

It would then be interesting to perform the calculations exemplified in this paper for such a case. Besides completing the literature on a specific class of supergravity duals of QCD-like (or Yang-Mills-like) theories, this paper sets the stage for potentially exciting future studies, in which the background of the supergravity dual already contains a geometric realisation of symmetry breaking with potential implications for model building. An example could be based upon the construction in~\cite{Karndumri:2012vh}, which includes the coupling to vector multiplets in the six-dimensional theory.
 
 \vspace{1cm}

\begin{acknowledgments}
The work of MP has been supported in part by the STFC Consolidated Grants ST/L000369/1 and ST/P00055X/1. JR is supported by STFC, through the studentship ST/R505158/1. DE is supported by the OCEVU Labex (ANR-11-LABX-0060) and the A*MIDEX project (ANR-11-IDEX-0001-02) funded by the ``Investissements d'Aveni'' French government program managed by the ANR. DE was also supported in part by the ERC Starting Grant HoloLHC-306605 and by the grant MDM-2014-0369 of ICCUB. MP would like to thank B.~Lucini and D.~C.~Thompson for useful discussions. 
\end{acknowledgments}

\appendix

\section{Massive vectors in $D=4$ dimensions}
\label{Sec:A}

We summarise in  this first Appendix some known results and conventions about the notation we adopt in $D=4$ dimensions, for the main purpose of keeping track of minus signs and factors of $2$, but also in order to facilitate direct comparison with the intermediate results outlined for the $D=5$ dimensional case.
 
All the individual relations exhibited in Appendix~\ref{Sec:A1} can be found in standard textbooks, but we find it useful to collect them all in one place, written with consistent conventions. Appendix~\ref{Sec:A2} provides an equivalent description of the same physics, as we generalise the analysis and results from~\cite{Bijnens:1995ii}, for the main purpose of exhibiting explicitly the role of gauge invariance in the different formulations of the same theory.

\subsection{About four-dimensional spontaneously broken $U(1)$ gauge theories}
\label{Sec:A1}

In $D=4$ dimensions, with space-time  signature $\{-\,,\,+\,,\,+\,,\,+\}$, a weakly coupled, spontaneously broken $U(1)$
gauge theory is described by the Lagrangian density
\beqs
{\cal L}_0&=&-\frac{1}{4}F_{\mu\nu}F^{\mu\nu} -\frac{1}{2}\left(\frac{}{}\partial_{\mu}\pi + m A_{\mu}\right)
\left(\frac{}{}\partial^{\mu}\pi + m A^{\mu}\right)\,,
\label{Eq:U(1)}
\eeqs
where $F_{\mu\nu}\equiv\partial_{\mu}A_{\nu}-\partial_{\nu}A_{\mu}$ is the field strength tensor of the vector field $A_{\mu}$, while $\pi$ is a pseudo-scalar field and $m$ the mass.
The $U(1)$ transformations are
\beqs
\pi&\rightarrow&\pi+ m \alpha\,,~~~~~~~~A_{\mu}\,\rightarrow\,A_{\mu}-\partial_{\mu}\alpha\,,
\eeqs
for $\alpha$ a generic function of the coordinates $x^{\mu}$. Both $F_{\mu\nu}$ and $\partial_{\mu}\pi + m A_{\mu}$ are gauge invariant for any value of $m$.

The customary quantisation procedure requires the introduction of the path integral that depends on source terms that we 
collectively and schematically denote by $J$, but do not write explicitly:
 \beqs
 {\cal Z}[J]&\equiv&{\cal N}_0 \int {\cal D} A_{\mu} {\cal D} \pi e^{i\int \di^4 x \,\left( {\cal L}_0+\, {\cal L}_{\rm g.f.}+{\rm sources}\right)}\,.
\label{Eq:gf}
 \eeqs
The gauge-fixing part of the Lagrangian is chosen to be
 \beqs
 {\cal L}_{\rm g.f.}&=&-\frac{1}{2\xi}\left(\partial^{\mu}A_{\mu}+\xi m \pi\right)\left(\partial^{\nu}A_{\nu}+\xi m \pi\right)\,,
 \eeqs
 so that the Lagrangian density becomes
 \beqs
 {\cal L}_0+\, {\cal L}_{\rm g.f.}
 &=&
 -\frac{1}{4}F_{\mu\nu}F^{\mu\nu}-\frac{1}{2}m^2A_{\mu}A^{\mu}-\frac{1}{2\xi}(\partial^{\mu}A_{\mu})^2\,+\nonumber\\
 &&-\frac{1}{2}\partial_{\mu}\pi\partial^{\mu}\pi-\frac{1}{2}\xi m^2 \pi^2 \,-\,\partial^{\mu}\left[\frac{}{}m \pi A_{\mu}\right]\,.
 \eeqs
The total derivative can be ignored, and the classical equations for vectors and \linebreak (pseudo-)scalars decouple from each other. One Fourier transforms to momentum-space, by making use of the following relations:\footnote{With this convention the  Fourier transform and its inverse have the same normalisation, in contrast with the more commonly used convention.}
\beqs
\psi(x^{\mu})&\equiv&\int \frac{\di^4q}{(2\pi)^2}e^{i q_{\mu}x^{\mu}}\tilde{\psi}(q_{\mu})\,,\\
\delta^{(4)}(q_{\mu})&\equiv&\int \frac{\di^4x}{(2\pi)^4}e^{i q_{\mu}x^{\mu}}\,.
\eeqs
We drop the $\tilde{}$ in the Fourier-transformed functions throughout the paper.

One then rewrites the functional ${\cal Z}[J]$, generator of all the correlation functions, as
 \beqs
  {\cal Z}[J]&\equiv&{\cal N}_0
  \int {\cal D} A_{\mu} {\cal D} \pi e^{i\int \di^4 q \,\left( \tilde{\cal L}_0+\, \tilde{\cal L}_{\rm g.f.}+{\rm sources}\right)}\,,
  \eeqs
  with
  \beqs
 \tilde{\cal L}_0+\, \tilde{\cal L}_{\rm g.f.}
 &=&
 -\frac{1}{2}A_{\mu}(-q)q^2P^{\mu\nu}A_{\nu}(q)-\frac{1}{2}m^2A_{\mu}(-q)\eta^{\mu\nu}A_{\nu}(q)
 -\frac{1}{2\xi}A_{\mu}(-q){q^{\mu}q^{\nu}}A_{\nu}(q)\,+\,\nonumber\\
 &&-\frac{1}{2}q^2\pi(-q)\pi(q)-\frac{1}{2}\xi m^2 \pi(-q)\pi(q)\,.
 \eeqs
In this expression, there appears the tensor
\beqs
P^{\mu\nu}(q^2)&\equiv&\eta^{\mu\nu}-\frac{q^{\mu}q^{\nu}}{q^2}\,,
\label{Eq:projector}
\eeqs
that obeys the transversality relation $q_{\mu}P^{\mu\nu}=0$.  The relations $P^{\mu\nu}+ \frac{q^{\mu}q^{\nu}}{q^2}=\eta^{\mu\nu}$,
and  $P^{\mu\nu}P_{\nu}^{\,\,\,\sigma}=P^{\mu\sigma}$,
imply  that  $P^{\mu\nu}$ and $\frac{q^{\mu}q^{\nu}}{q^2}$ are, respectively, the projectors on the transverse and longitudinal polarisations.

The propagator for the vectors in the general $R_{\xi}$-gauge reads
\beqs
(D_F)^{\mu\nu}
&=&
 \frac{-i}{q^2+m^2}\left(\eta^{\mu\nu}-\frac{q^{\mu}q^{\nu}}{q^2}\right)+
 \frac{-i}{q^2/\xi+m^2}\frac{q^{\mu}q^{\nu}}{q^2}\,,
\eeqs
and satisfies the equation
\beqs
i\delta_{\mu}^{\,\,\,\rho}
&=&
\left[-\left(\eta^{\mu\nu}-\frac{q^{\mu}q^{\nu}}{q^2}\right)\left(q^2+m^2\right)-\frac{q^{\mu}q^{\nu}}{q^2}\left(q^2+\xi m^2\right)\frac{1}{\xi}\right](D_F)_\nu{}^\rho\,.
\eeqs
The propagator for the (would-be) Goldstone boson $\pi$ is
\beqs
D_{\pi}&=&\frac{-i}{q^2+\xi m^2}\,,
\eeqs
where the numerator is $-i$ rather than $i$ because of the signature $\{-,+,+,+\}$. The propagator of the longitudinal part of the vectors can be obtained from the one of the transverse parts by replacing $q^2\rightarrow q^2/\xi$, and furthermore its poles coincide with those of the Goldstone propagator, for any $\xi$. Only the transverse part of the vector propagator is $\xi$-independent.

In textbooks, the choice $\xi=1$ is referred to as Feynman gauge, in which the propagator of the vectors is proportional to $\eta_{\mu\nu}$, while $\xi=0$ is the Landau gauge, in which all vectors are transverse. The unitary gauge is obtained by setting $\xi\rightarrow +\infty$, so that only physical degrees of freedom remain.

\subsection{2-forms in $D=4$ dimensions}
\label{Sec:A2}

In $D=4$ dimensions, a massless 2-form is equivalent to a massless 0-form (a scalar), while a massive 2-form is equivalent to a massive 1-form (a vector). We follow closely the discussion in Ref.~\cite{Bijnens:1995ii} (see also~\cite{Ecker:1989yg,Bruns:2004tj}), and generalise it in this Appendix to show the equivalence explicitly, by highlighting both the role of the gauge redundancies in the various formulations of the same theory, and also the peculiarities of the four-dimensional case. We conclude by briefly mentioning some of the subtleties appearing  in higher dimensions. We assume the metric(s) to be flat, and we restrict attention to the $U(1)$ theory. In the non-abelian case some partial derivatives  have to be generalised to covariant derivatives, the field-strengths transform as tensors, rather than being invariant, and furthermore, one has to keep track of the Faddeev-Popov ghosts, none of which significantly affect the results.

We start from Eq.~(\ref{Eq:U(1)}) in Appendix~\ref{Sec:A1}, 
by defining the 2-forms $B_{\mu\nu}$ and $\tilde{B}_{\mu\nu}$:
\beqs
\label{Eq:replacements}
\partial_{\mu} \pi + m A_{\mu} &\equiv& \frac{1}{2}\epsilon_{\mu\nu\rho\sigma} \partial^{\nu}B^{\rho\sigma}\,\equiv\,\partial^{\nu}\tilde{B}_{\mu\nu}\,.
\eeqs
These definitions introduce a gauge invariance, as we could replace
\beqs
\label{eq:vectorgaugeinvariance}
B_{\mu\nu}&\rightarrow&B_{\mu\nu}-{2}\partial_{[\mu}\alpha_{\nu]}\,,
\eeqs
with $\alpha_{\mu}$ a vectorial function of the coordinates $x^{\mu}$, without affecting either $\tilde{B}_{\mu\nu}$ or the combinations $\partial_{\mu} \pi + m A_{\mu}$. We will return to this point later on.

The trivial functional identities
\beqs
1&=&{\cal N}_1\int {\cal D} {\cal I}^{\prime}_{\mu\nu}  e^{i\int \di^4 x \, {\cal I}^{\prime}_{\mu\nu}{\cal I}^{\prime\,\mu\nu}}\,,\\
1&=&
\int {\cal D} \tilde{B}_{\mu\nu} \left|{\rm det}\left( \frac{\partial}{m}\right)\right| 
\delta\left(\frac{}{}A_{\mu}+\frac{1}{m}\partial_{\mu}\pi-\frac{1}{m}\partial^{\nu}\tilde{B}_{\mu\nu}\right)\\
&=& \nonumber
{\cal N}_2\int {\cal D} \tilde{B}_{\mu\nu} \,
\delta\left(\frac{}{}A_{\mu}+\frac{1}{m}\partial_{\mu}\pi-\frac{1}{m}\partial^{\nu}\tilde{B}_{\mu\nu}\right)\,,
\eeqs
allow for the rewriting of Eq.~(\ref{Eq:gf}) as follows:
\beqs
{\cal Z}[J]&=&{\cal N} \int {\cal D} A_{\mu} {\cal D} \pi \nonumber
{\cal D} {\cal I}^{\prime}_{\mu\nu}  
{\cal D}\tilde{B}_{\mu\nu} \delta\left(\frac{}{}A_{\mu}+\frac{1}{m}\partial_{\mu}\pi-\frac{1}{m}\partial^{\nu}\tilde{B}_{\mu\nu}\right)\times\\
&&
\times\,\,  e^{i\int \di^4 x \, {\cal I}^{\prime}_{\mu\nu}{\cal I}^{\prime\,\mu\nu}}\,e^{i\int \di^4 x 
\,\left( {\cal L}_0+{\cal L}_{\rm g.f.}+{\rm sources}\right)}\,,
\eeqs
where  ${\cal N}={\cal N}_0{\cal N}_1{\cal N}_2$ is a $J$-independent (divergent) normalisation. 
 
By performing the integral over $A_{\mu}$---hence making use of the $\delta$-function to replace $A_{\mu}$---the dependence on $\pi$ disappears from ${\cal L}_0(\pi,A_{\mu})={\cal L}_0(\tilde{B}_{\mu\nu})$:
\beqs
\nonumber
{\cal L}_0(\tilde{B}_{\mu\nu})&=&-\frac{1}{4m^2}\left[\frac{}{}\partial_{\mu}\partial^{\alpha}\tilde{B}_{\nu\alpha}
-\partial_{\nu}\partial^{{\alpha}}\tilde{B}_{\mu\alpha}\right]
\left[\frac{}{}\partial^{\mu}\partial^{\bar{\alpha}}\tilde{B}^{\nu}_{\,\,\,\,\bar{\alpha}}
-\partial^{\nu}\partial^{\bar{\alpha}}\tilde{B}^{\mu}_{\,\,\,\,\bar{\alpha}}\right]
-\frac{1}{2}\partial^{\alpha}\tilde{B}_{\mu\alpha}\partial^{\bar{\alpha}}\tilde{B}^{\mu}_{\,\,\,\,\bar{\alpha}}
\label{Eq:L0Bmunu}\\
&\equiv& -F_{\mu\nu}[\tilde{B}]^2-\frac{1}{2}\partial^{\alpha}\tilde{B}_{\mu\alpha}\partial^{\bar{\alpha}}\tilde{B}^{\mu}_{\,\,\,\,\bar{\alpha}}\,.
\eeqs 
This is a consequence of the fact that  ${\cal L}_0$ depends only on the gauge-invariant combinations $\partial_{\mu}\pi+m A_{\mu}$ and $F_{\mu\nu}$. Conversely, in the gauge-fixing part, because $\tilde{B}_{\mu\nu}$ is antisymmetric, one sees that $\partial^{\mu}\partial^{\nu}\tilde{B}_{\mu\nu}=0$, and hence one finds that ${\cal L}_{\rm g.f.}(\pi,A_{\mu})={\cal L}_{\rm g.f.}(\pi)$:
\beqs
\label{eq:Lgfpi}
{\cal L}_{\rm g.f.}(\pi)&=&-\frac{1}{2\xi}\left(-\frac{1}{m}\partial^{\mu}\partial_{\mu}\pi+\xi m \pi\right)^2\,.
\eeqs
The equation of motion derived from Eq.~\eqref{eq:Lgfpi} reproduces the on-shell condition for the \linebreak (pseudo-)scalar field $\pi$---which in Fourier space reads $(q^2+\xi m^2)\pi=0$. The integral over $\pi$ amounts to another redefinition of the overall normalisation constant ${\cal N}^{\prime}={\cal N}\int {\cal D}\pi e^{i\int\di^4x {\cal L}_{\rm g.f.}}$, and one finds
\beqs
{\cal Z}[J]&=&{\cal N}^{\prime} \int 
{\cal D} {\cal I}^{\prime}_{\mu\nu}  
{\cal D}\tilde{B}_{\mu\nu} 
e^{i\int \di^4 x \,\left( {\cal I}^{\prime}_{\mu\nu}{\cal I}^{\prime\,\mu\nu} +{\cal L}_0+{\rm sources}\right)}\,.
\eeqs

The difficulty at this point is represented by the appearance of kinetic terms with four derivatives  in Eq.~(\ref{Eq:L0Bmunu}), which superficially would lead to potential violations of causality. Following~\cite{Bijnens:1995ii}, one performs the change of variable
\beqs
{\cal I}_{\mu\nu}^{\prime}&\equiv&\hat{\mu} {\cal I}_{\mu\nu} +F_{\mu\nu}(\tilde{B})\,,
\eeqs
and arrives at
\beqs
{\cal Z}[J]&=&{\cal N}^{\prime}\int {\cal D}\tilde{B}_{\mu\nu} {\cal D} (\hat{\mu} {\cal I}_{\mu\nu})
e^{i\int \di^4 x \, {\cal L}_I+{\rm sources}}\,,
\eeqs
where the $F_{\mu\nu}[\tilde{B}]^2$ term cancelled,  while the Lagrangian density is (up to a total derivative)
\beqs
{\cal L}_I&=&-\frac{1}{2}\partial^{\nu}\tilde{B}_{\mu\nu}\partial^{{\rho}}\tilde{B}^{\mu}_{\,\,\,\,{\rho}}
+\hat{\mu}^2 {\cal I}_{\mu\nu} {\cal I}^{\mu\nu}
+\frac{2 \hat{\mu}}{m}\partial^{\nu}{\cal I}_{\mu\nu}\partial^{{\rho}}{\tilde{B}}^{\mu}_{\,\,\,\,{\rho}}\,.
\eeqs 
Notice that $\hat{\mu}$ has dimension of a mass, as do the two 2-forms ${\cal I}_{\mu\nu}$ and ${\tilde{B}}_{\mu\nu}$. The four-derivative term has been traded for the doubling of the tensor-field content. One then diagonalises the system of tensors, a process that we show in detail.
 
After rotating according to
\beqs
\tilde{B}_{\mu\nu}&=&\cos \theta\, {G}_{\mu\nu}+\sin\theta\, H_{\mu\nu}\,,\\
{\cal I}_{\mu\nu}&=&- \sin \theta \,{G}_{\mu\nu}+\cos\theta\, H_{\mu\nu}\,,
\eeqs
with $\tan (2\theta) =\frac{4\hat{\mu}}{m}$\,,
and then rescaling the resulting fields according to
\beqs
\tilde{G}_{\mu\nu}&=& \frac{\cos(\theta)}{\sqrt{\cos(2\theta)}} G_{\mu\nu}\,,\\
\tilde{H}_{\mu\nu}&=& - \frac{\sin(\theta)}{\sqrt{\cos(2\theta)}} H_{\mu\nu}\,,
\eeqs
the resulting Lagrangian density is given by
\beqs
{\cal L}_I&=&-\frac{1}{2}\partial^{\nu}\tilde{G}_{\mu\nu}\partial^{\rho}\tilde{G}^{\mu}_{\,\,\,\,\rho}
+\frac{1}{2}\partial^{\nu}\tilde{H}_{\mu\nu}\partial^{\rho}\tilde{H}^{\mu}_{\,\,\,\,\rho}+\\
&& \nonumber
+\hat{\mu}^2\cos(2\theta)\left(\tilde{G}_{\mu\nu}\,,\,\tilde{H}_{\mu\nu}\right)
\left(\begin{array}{cc}
\tan^2\theta & 1 \cr
1 & \frac{1}{\tan^2\theta}
\end{array}\right)
\left(\begin{array}{c}
\tilde{G}^{\mu\nu} \cr \tilde{H}^{\mu\nu}\end{array}\right)\,.
\eeqs
The  2-forms have kinetic terms with opposite signs. With our choice of metric signature, the kinetic term of $\tilde{H}_{\mu\nu}$ is compatible with causal propagation.

The final step consists of diagonalizing the mass terms. Because the kinetic term is the matrix ${\rm diag} \left(-1\,,\,1\right)$,
the transformation involves hyperbolic functions
\beqs
\tilde{G}_{\mu\nu}&=&\cosh\beta\,W_{\mu\nu}+\sinh\beta\,K_{\mu\nu}\,,\\
\tilde{H}_{\mu\nu}&=&\sinh\beta\,W_{\mu\nu}+\cosh\beta\,K_{\mu\nu}\,,
\eeqs
and the condition for the mass term to be diagonal is satisfied by demanding that 
\beqs
\beta&=&\frac{1}{2}\log (\cos (2\theta))\,,
\eeqs
so that finally the Langragian density is given by 
\beqs
{\cal L}_I&=&\frac{1}{2}\partial^{\alpha}{K}_{\mu\alpha}\partial^{\bar{\alpha}}{K}^{\mu}_{\,\,\,\,\bar{\alpha}}
-\frac{1}{2}\partial^{\alpha}{W}_{\mu\alpha}\partial^{\bar{\alpha}}{W}^{\mu}_{\,\,\,\,\bar{\alpha}}
+\frac{m^2}{4}K_{\mu\nu}K^{\mu\nu}\,.
\eeqs
At this point, the parameter $\hat{\mu}$ has disappeared, and there is no mixing present between $K_{\mu\nu}$ and $W_{\mu\nu}$. The latter is unstable, but only provides another factorised contribution to the normalisation of the path integral,
barring some subtleties in the definition of the sources that we do not report here (but see~\cite{Bijnens:1995ii}). The path integral is then
\beqs
{\cal Z}[J]&=&{\cal N}^{\prime\prime}\int {\cal D} K_{\mu\nu} e^{i\int \di^4 x {\cal L}_K \,+\, {\rm sources}}\,,
\eeqs
where the action of the massive 2-form $K_{\mu\nu}$, with mass $m$ is
\beqs
\label{eq:LagK}
{\cal L}_K&=&\frac{1}{2}\partial^{\alpha}{K}_{\mu\alpha}\partial^{\bar{\alpha}}{K}^{\mu}_{\,\,\,\,\bar{\alpha}}
+\frac{m^2}{4}K_{\mu\nu}K^{\mu\nu}\,.
\label{Eq:dualtensor}
\eeqs

We can go back now and reconstruct the analogue of the second identity in Eq.~(\ref{Eq:replacements}), by further redefining the 2-form ${\cal B}_{\mu\nu}$ via the relation:
\beqs
K_{\mu\nu}&\equiv&\frac{1}{2}\epsilon_{\mu\nu\rho\sigma}\left(\frac{}{}{\cal B}^{\rho\sigma}+\frac{1}{m}{\cal F}^{\rho\sigma}\right)
\,\equiv\,\frac{1}{2m}\epsilon_{\mu\nu\rho\sigma}{\cal H}^{\rho\sigma}
\,,
\eeqs
with
\beqs
{\cal F}^{\rho\sigma}&\equiv&\partial^{\rho}{\cal A}^{\sigma}-\partial^{\sigma}{\cal A}^{\rho}\,,
\eeqs
and ${\cal A}_{\mu}$ an Abelian gauge field. Because ${\cal F}$ is exact, it is also closed, and hence
\beqs
\partial^{\alpha}K_{\mu\alpha}&=&\frac{1}{2}\epsilon_{\mu\alpha\rho\sigma}\partial^{\alpha}{\cal B}^{\rho\sigma}
\eeqs
is independent of ${\cal A}_{\mu}$. As anticipated just after Eq.~(\ref{Eq:replacements}), this is a manifestation of the fact that ${\cal B}^{\mu\nu}$ is defined up to a gauge transformation with parameter(s) ${\cal A}_{\mu}$, which extends the original $U(1)$ gauge invariance of Eq.~(\ref{Eq:U(1)}).

By making use of the identities
\beqs
\epsilon_{\mu\nu\rho\sigma}\epsilon^{\mu\nu}_{\,\,\,\,\,\,\,\,\bar{\rho}\bar{\sigma}}&=&
-{2}\left(\frac{}{}\eta_{\rho\bar{\rho}}\eta_{\sigma\bar{\sigma}}-\eta_{\rho\bar{\sigma}}\eta_{\sigma\bar{\rho}}\right)\,,\\
\epsilon_{\mu\nu\rho\sigma}\epsilon^{\mu}_{\,\,\,\,\bar{\nu}\bar{\rho}\bar{\sigma}}&=&
-\frac{}{}\eta_{\nu\bar{\nu}}\eta_{\rho\bar{\rho}}\eta_{\sigma\bar{\sigma}}
+\eta_{\nu\bar{\nu}}\eta_{\rho\bar{\sigma}}\eta_{\sigma\bar{\rho}}
+\frac{}{}\eta_{\nu\bar{\rho}}\eta_{\rho\bar{\nu}}\eta_{\sigma\bar{\sigma}}+\\
&&\nonumber
-\eta_{\nu\bar{\rho}}\eta_{\rho\bar{\sigma}}\eta_{\sigma\bar{\nu}}
-\frac{}{}\eta_{\nu\bar{\sigma}}\eta_{\rho\bar{\nu}}\eta_{\sigma\bar{\rho}}
+\eta_{\nu\bar{\sigma}}\eta_{\rho\bar{\rho}}\eta_{\sigma\bar{\nu}}
\,,
\eeqs
Eq.~\eqref{eq:LagK} can be rewritten by trading $K_{\mu\nu}$ for ${\cal B}_{\mu\nu}$ and ${\cal F}_{\mu\nu}$, so that
\beqs
{\cal L}_K&=&-\frac{1}{12}{G}_{\mu\nu\rho}G^{\mu\nu\rho}
-\frac{1}{4}{\cal H}^{\mu\nu}{\cal H}_{\mu\nu}\,.
\label{Eq:tensor}
\eeqs
This Lagrangian in $D=4$ dimensions for flat space and sigma-model metrics is adopted in higher dimensions (see for example Eq.~(\ref{Eq:2form}), later in Appendix~\ref{Sec:B3}), to describe $2$-forms fields. The field-strength ${G}_{\mu\nu\rho}$ of ${\cal B}_{\mu\nu}$ is completely anti-symmetrised:
\beqs
{G}_{\mu\nu\rho}&=&3\partial_{[\mu}{\cal B}_{\nu\rho]}\,=\,\partial_{\mu}{\cal B}_{\nu\rho}+
\partial_{\rho}{\cal B}_{\mu\nu}+
\partial_{\nu}{\cal B}_{\rho\mu}\,.
\eeqs
The Lagrangian density  for $m=0$ consists of the simple kinetic term for a massless scalar (dual to ${\cal B}_{\mu\nu}$) and a massless $U(1)$ gauge boson ${\cal A}_{\mu}$, while the coupling in the mass term reinstates gauge invariance when $m\neq 0$. Quantisation then requires the integration over both ${\cal A}_{\mu}$ and ${\cal B}_{\mu\nu}$, and to introduce appropriate gauge-fixing terms.

Summarising, in $D=4$ dimensions, one can use equivalently any of the three Lagrangian densities in Eq.~(\ref{Eq:U(1)}), or Eq.~(\ref{Eq:dualtensor}) or Eq.~(\ref{Eq:tensor}), and describe exactly the same physics. The three have different gauge symmetries: there is no invariance in Eq.~(\ref{Eq:dualtensor}), while Eq.~(\ref{Eq:U(1)}) is invariant under a $U(1)$ transformation with a parameter $\alpha$ and Eq.~(\ref{Eq:tensor}) has a gauge invariance parametrised by a vector as in Eq.~\eqref{eq:vectorgaugeinvariance}. As such, quantisation requires different path integrals and different gauge-fixing terms. In particular, it is usually convenient in $D=4$ to use Eq.~(\ref{Eq:U(1)}), so that one has to write only 0-forms and 1-forms, while ignoring higher-order forms.

In higher dimensions, one might try to generalise this line of argument. For example, if the number of dimensions is $D=2p+1$, a massive $p$-form can be written in terms of two massless $p$-forms soldered together by a first-order differential operator that introduces the mass term (see for example~\cite{Noronha:2003vp} for a nicely pedagogical discussion). But  the equations of motion of the $p$-forms involve their duals, because the mass terms are written with the completely anti-symmetric tensor $\epsilon^{\mu_1\cdots\mu_D}$.

For example, in $D=5$ dimensions a massless 2-form is dual to a massless 1-form (each propagating 3 physical degrees of freedom). It is tempting to think of the massive 2-form (6 degrees of freedom) as the result of soldering the two massless 2-forms 
dual to two massless 1-forms, and hence try to write the Lagrangian just in terms of vectors. But the soldering term requires the dualisation of one of the forms, and hence the result is that we must either keep track in the action of  bilinear terms involving both the forms and their duals, or write the theory in terms of one 1-form and one 2-form, and then apply the Higgs mechanism,
which is what we do in this paper, as is shown explicitly in Appendix~\ref{Sec:B3}. In the broad context of gauged supergravities, for which we refer the reader to~\cite{Samtleben:2008pe} and references therein, similar considerations are in fact enforced on rather general grounds.

\section{Bosonic fields in $D=5$ dimensions}
\label{Sec:B}

In this Appendix, we collect general results about the treatment of scalar and $p$-form fields coupled to gravity in $D=5$ dimensions, of relevance to this paper. We emphasise the role of gauge invariance in the discussion of the fluctuations on a given sigma-model background coupled to gravity. Appendix~\ref{Sec:B1} contains the treatment of the sigma-model scalar and tensor fluctuations, while in Appendix~\ref{Sec:B2} and Appendix~\ref{Sec:B3} we treat $p$-forms.

The separate treatment of these sectors hinges on the assumption that only the metric and the sigma-model scalar fields acquire non-trivial profiles in the bulk, and the fact that one only needs to retain terms up to second order in the fluctuations in order to compute spectra.  As a result, the treatment of diffeomorphism invariance can be performed independently (in Appendix~\ref{Sec:B1}) from that of the gauge invariance inherent in the formulation of theories with $p$-forms (in Appendix~\ref{Sec:B2} and~\ref{Sec:B3}). We will form gauge-invariant combinations of the various fluctuations, the equations of motion and boundary conditions for which will give us the spectrum.

\subsection{About sigma-models coupled to gravity in $D=5$ dimensions}
\label{Sec:B1}

We start from the conventions we adopt for gravity. The Christoffel symbol is
\beqs
\Gamma^P_{\,\,\,\,MN}&\equiv&\frac{1}{2}g^{PQ}\left(\frac{}{}\partial_Mg_{NQ}+\partial_Ng_{QM}-\partial_Qg_{MN}\right)\,,
\eeqs
the Riemann tensor is
\beqs
R_{MNP}^{\,\,\,\,\,\,\,\,\,\,\,\,\,\,\,\,\,Q}&\equiv&\partial_N\Gamma^Q_{\,\,\,\,MP}-\partial_M\Gamma^Q_{\,\,\,\,NP}+\Gamma^S_{\,\,\,\,MP}\Gamma^Q_{\,\,\,\,SN}
-\Gamma^S_{\,\,\,\,NP}\Gamma^Q_{\,\,\,\,SM}\,,
\eeqs
the Ricci tensor is
\beqs
R_{MN}&\equiv&R_{MPN}^{\,\,\,\,\,\,\,\,\,\,\,\,\,\,\,\,\,P}\,,
\eeqs
and finally the Ricci scalar is
\beqs
R&\equiv&R_{MN}g^{MN}\,.
\eeqs
The covariant derivative with respect to gravity for a $(1,1)$-tensor takes the form
\beqs
\nabla_M T^{P}_{\,\,\,\,N}&\equiv&\partial_MT^{P}_{\,\,\,\,N}+\Gamma^P_{\,\,\,\,MQ}T^{Q}_{\,\,\,\,N}-\Gamma^Q_{\,\,\,\,MN}T^{P}_{\,\,\,\,Q}\,,
\eeqs
and can be generalised to any tensor.

Much in the same way,  the sigma-model connection descends from the sigma-model metric $G_{ab}$---with $a,b=1\,,\,\cdots\,,\,n$ the indexes in the $n$-dimensional scalar manifold---as follows
\beqs
{\cal G}^d_{\,\,\,\,ab}&\equiv& \frac{1}{2}G^{dc}\left(\frac{}{}\partial_aG_{cb}+\partial_bG_{ca}-\partial_cG_{ab}\right)\,.
\eeqs
The sigma-model Riemann tensor is\footnote{Notice that the only difference in the conventions for the two Riemann tensors is the reversed ordering in which one writes  the indexes.}
\beqs
{\cal R}^a_{\,\,\,\,bcd}
&\equiv& \partial_c{\cal G}^a_{\,\,\,\,bd}-\partial_d{\cal G}^a_{\,\,\,\,bc}+{\cal G}^e_{\,\,\,\,bd}{\cal G}^a_{\,\,\,\,ce}-{\cal G}^e_{\,\,\,\,bc}{\cal G}^a_{\,\,\,\,de}\,,
\eeqs
while the sigma-model covariant derivative is
\beqs
D_b X^d_{\,\,\,\,a}&\equiv& \partial_b X^d_{\,\,\,\,a}+{\cal G}^d_{\,\,\,\,cb}X^c_{\,\,\,\,a}-{\cal G}^c_{\,\,\,\,ab}X^d_{\,\,\,\,c}\,,
\eeqs
in terms of the sigma-model derivative  $\partial_b=\frac{\partial}{\partial \Phi^b}$.

The space being bounded, and consisting of a five-dimensional manifold and two (four-dimensional) boundaries, we need the  induced metric, which  is given by
\beqs
\tilde{g}_{MN}&\equiv& g_{MN}-N_MN_M\,,
\eeqs
in terms of the vector $N_M$ ortho-normal to the boundary, and satisfies the defining properties:
\beqs
g_{MN}N^MN^N&=&1\,,~~~~~~~~\tilde{g}_{MN} N^N\,=\,0\,.
\eeqs
The vector $N_M$ is oriented to point outwards from the boundary. The extrinsic curvature is computed in terms of the symmetric tensor
\beqs
K_{MN}&\equiv&\nabla_MN_N\,=\,\partial_M N_N-\Gamma^Q_{\,\,\,\,MN}N_Q
\eeqs
and is given by $K\equiv \tilde{g}^{MN}K_{MN}$.

The action in $D=5$ dimensions is then written to agree with the conventions in~\cite{Elander:2010wd}:
\beqs
\label{Eq:Sigma}
{\cal S}_5 &=& \int \di^4 x \di r\left\{ \sqrt{-g_5}\left[\frac{R}{4} +{\cal L}_5\right] +\sum_{i=1,2}\delta(r-r_i)(-)^i\sqrt{-g_5}\left[\frac{K}{2}+{\cal L}_i\right]\right\}\,.
\eeqs
The matter Lagrangian density in the bulk is given by
\beqs
{\cal L}_5 &=& -\frac{1}{2}G_{ab}g^{MN} \partial_M \Phi^a \partial_N \Phi^b - \mathcal V(\Phi^a)\,,
\eeqs
while ${\cal L}_i$ are boundary-localised contributions to the scalar part of the action.

We begin by reviewing the gauge-invariant formalism developed in \cite{Bianchi:2003ug,Berg:2006xy,Berg:2005pd,Elander:2009bm,Elander:2010wd} (to which the reader is referred for details) that allows for the computation of the scalar and tensorial parts of the spectrum. We start by expanding the scalar fields as
\beq
	\Phi^a(x^\mu,r) = \bar \Phi^a(r) + \varphi^a(x^\mu,r) \,,
\eeq
where $\varphi^a(x^\mu,r)$ are small fluctuations around the background solution $\bar \Phi^a(r)$. Decomposing the metric according to the Arnowitt-Deser-Misner (ADM) formalism \cite{Arnowitt:1959ah}, we write
\beqs
	\dd s_5^2 &=& \left( (1 + \nu)^2 + \nu_\sigma \nu^\sigma \right) \dd r^2 + 2 \nu_\mu \dd x^\mu \dd r + e^{2A(r)} \left( \eta_{\mu\nu} + h_{\mu\nu} \right) \dd x^\mu \dd x^\nu \,, \\
	h^\mu{}_\nu &=& \mathfrak e^\mu{}_\nu + i q^\mu \epsilon_\nu + i q_\nu \epsilon^\mu + \frac{q^\mu q_\nu}{q^2} H + \frac{1}{3} \delta^\mu{}_\nu h,
\eeqs
where $\mathfrak e^\mu{}_\nu$ is transverse and traceless, $\epsilon^\mu$ is transverse, and the four-dimensional indices $\mu$, $\nu$ are raised and lowered by the boundary metric $\eta$. We treat $\nu(x^\mu,r)$, $\nu^\mu(x^\mu,r)$, $\mathfrak e^\mu{}_\nu(x^\mu,r)$, $\epsilon^\mu(x^\mu,r)$, $H(x^\mu,r)$, and $h(x^\mu,r)$ as small fluctuations around the background metric determined by the warp factor $A(r)$.

Under infinitesimal diffeomorphisms $\xi^M(x^\mu,r)$, the fluctuations transform as
\beqs
	\delta \varphi^a &=& \partial_r \bar \Phi^a \xi^r \,, \ \ \ \delta \nu = \partial_r \xi^r \,, \ \ \ \delta H = 2 \partial_\mu \xi^\mu \,, \ \ \ \delta h = 6 \partial_r A \xi^r \,, \\ \delta \nu^\mu &=& \partial^\mu \xi^r \,, \ \ \ \delta \epsilon^\mu = P^\mu{}_\nu \xi^\nu \,, \ \ \ \delta \mathfrak e^\mu{}_\nu = 0 \,,
\eeqs
where we have neglected terms higher than linear order in the fluctuations themselves. After forming the gauge-invariant (under diffeomorphisms) combinations (in addition to the gauge invariant variable $\mathfrak e^\mu{}_\nu$)
\beqs
	\mathfrak a^a &=& \varphi^a - \frac{\partial_r \bar \Phi^a}{6\partial_r A} h \,, \\
	\mathfrak b &=& \nu - \partial_r \left( \frac{h}{6\partial_r A} \right) \,, \\
	\mathfrak c &=& e^{-2A} \partial_\mu \nu^\mu - \frac{e^{-2A} q^2 h}{6 \partial_r A} - \frac{1}{2} \partial_r H \,, \\
	\mathfrak d^\mu &=& e^{-2A} P^\mu{}_\nu \nu^\nu - \partial_r \epsilon^\mu \,,
\eeqs
the linearized equations of motion decouple into different sectors according to spin. For the tensorial fluctuations $\mathfrak e^\mu{}_\nu$, one obtains the equation of motion
\beq
\label{eq:tensoreom}
	\left[ \partial_r^2 + 4 \partial_r A \partial_r - e^{-2A(r)} q^2 \right] \mathfrak e^\mu{}_\nu = 0 \,,
\eeq
and boundary condition
\beq
\label{eq:tensorbc}
	\partial_r \mathfrak e ^\mu{}_\nu \big|_{r_i} = 0 \, .
\eeq
Together, Eqs.~(\ref{eq:tensoreom} - \ref{eq:tensorbc}) allow one to compute the tensor part of the spectrum. The equation of motion for $\mathfrak d^\mu$ is algebraic, and hence does not lead to a spectrum of composite states. The equations of motion for $\mathfrak b$ and $\mathfrak c$ are also algebraic, and can be solved in terms of $\mathfrak a^a$. Using this, the equations of motion for the scalar fluctuations can be written as
\beqs
\label{eq:scalareom}
	0 &=& \Big[ {\cal D}_r^2 + 4 \partial_{r}A {\cal D}_r - e^{-2A} q^2 \Big] \mathfrak{a}^a \,+\,\\ \nonumber
	&& - \Big[ \mathcal V^a{}_{|c} - \mathcal{R}^a{}_{bcd} \partial_{r}\bar\Phi^b \partial_{r}\bar\Phi^d + \frac{4 (\partial_{r}\bar\Phi^a \mathcal V^b + \mathcal V^a 
	\partial_{r}\bar\Phi^b) G_{bc}}{3 \partial_{r} A} + \frac{16 \mathcal V \partial_{r}\bar\Phi^a \partial_{r}\bar\Phi^b G_{bc}}{9 (\partial_{r}A)^2} \Big] \mathfrak{a}^c\,,
\eeqs
and the boundary conditions as
\beqs
\label{eq:scalarbc}
 \frac{2  e^{2A}\partial_{r}\bar \Phi^a }{3q^2 \partial_{r}A}
	\left[ \partial_{r}\bar \Phi^b{\cal D}_r -\frac{4 \mathcal V \partial_{r}\bar \Phi^b}{3 \partial_rA} - \mathcal V^b \right] \mathfrak a_b - \mathfrak a^a\Big|_{r_i} = 0 \, .
\eeqs
Here, $\mathcal V^a{}_{|b} \equiv \frac{\partial \mathcal V^a}{\partial \Phi^b} + \mathcal G^a_{\ bc} \mathcal V^c$, and the background covariant derivative is defined as $\mathcal D_r \mathfrak a^a \equiv \partial_r \mathfrak a^a + \mathcal G^a_{\ bc} \partial_r \bar \Phi^b \mathfrak a^c$. Eqs.~(\ref{eq:scalareom}) and~(\ref{eq:scalarbc}) allow us to compute the scalar part of the spectrum. 

Let us make a couple of comments about the boundary conditions for the tensors and scalars, reported in Eq.~\eqref{eq:tensorbc} and Eq.~\eqref{eq:scalarbc}. In order to make the variational problem well-defined, one introduces boundary-localised actions, consisting of the Gibbons-Hawking term for gravity, as well as an action for the scalar fields that is fixed by consistency requirements up to a term that is second order in the fluctuations. Taking the latter to be a boundary mass term for the fluctuations of the scalars, in the limit of infinite mass, one obtains the boundary condition $\varphi^a = 0$, which becomes Eq.~\eqref{eq:scalarbc} when written in terms of the gauge-invariant variable $\mathfrak a^a$. This boundary condition ensures that the subleading modes are retained, as the IR (UV) cutoffs are taken towards the end-of-space (boundary), in agreement with the standard prescription in gauge-gravity duality. The same is true for the tensorial modes when Eq.~\eqref{eq:tensorbc} is imposed in the IR (UV).

In order to calculate the renormalised two-point function, and obtain the spectrum from the location of its poles, a complete treatment making use of holographic renormalisation is necessary. It is possible to make the argument that the prescription outlined above captures the correct location of the poles, at least for $M^2 = - q^2 > 0$. The counter-terms are provided by a boundary action that is a functional of the boundary values of the bulk fields and derivatives thereof with respect to the boundary coordinates. Correlation functions are computed by differentiating with respect to the boundary values of the fields, and taking the limit of the UV cutoff $r_2 \rightarrow \infty$. The contribution of the counter-terms to the finite part of the renormalised two-point function is hence a polynomial function, and does not shift the location of the poles.

\subsection{Vectors in $D=5$ dimensions}
\label{Sec:B2}

A $U(1)$ theory in $D=5$ dimensions can be described by supplementing the sigma-model coupled to gravity by the following action:
\beqs
{\cal S}_5^{(1)} &=&\int \di^4x\di r \sqrt{-g_5}\left\{ 
-\frac{1}{4}H \,F_{MN}F_{RS}\,g^{MR}\,g^{NS} +\right.\\
&&\left.\nonumber-\frac{1}{2}G \left(\frac{}{}\partial_M \pi+m A_M\right)g^{MN}\left(\frac{}{}\partial_N \pi+m A_N\right)
\right\}\,,
\eeqs
where $G$ and $H$ are the sigma-model geometric factors, and depend on the background scalars $\Phi^a$, while $m$ is a symmetry-breaking parameter, and $F_{MN}\equiv 2\partial_{[M}A_{N]}=\partial_MA_N-\partial_NA_M$. The vector (1-form) $A_{M}$ and (pseudo-)scalar (0-form) $\pi$ obey the $U(1)$ transformation rules:
\beqs
\pi\,\rightarrow\,\pi+m \alpha\,,~~~~~~~~~~A_M\,\rightarrow\,A_M- \partial_M \alpha\,,
\eeqs
where $\alpha$ is a function of the space-time coordinates.

We decompose the fields in terms of four-dimensional quantities, in analogy with what is done in the ADM formalism applied to gravity. The fields $A_5$ and $\pi$ both behave as Goldstone bosons, the former as a consequence of the Kaluza-Klein decomposition, the latter in connection with the breaking of the $U(1)$ in $D=5$ dimensions. A combination of the two provides the longitudinal components for the infinite tower of massive vector states. Another combination  remains  in the spectrum, as a whole tower of massive pseudo-scalar particles.

After some algebra, in particular after Fourier-transforming in four dimensions, and performing some integrations by parts, we can rewrite the action as follows
\beqs
{\cal S}_5^{(1)} &=&\int \di^4q\di r \left\{ 
-\frac{1}{2}H\,A_{\mu}(-q)\,q^2 P^{\mu\nu} A_{\nu}(q)\,
-\frac{1}{2}H e^{2A} q^2 A_{5}(-q)A_5(q)
\,\right.\nonumber\\
&&\nonumber\left.
-\frac{1}{2}A_{\mu}(-q)\eta^{\mu\nu}\left[-\partial_r\left(H e^{2A} \partial_r A_{\nu}(q)\right)\right]
\,\right.\\
&&\nonumber\left.
+\sum_{i=1,2}(-)^i\delta(r-r_i)\left[-\frac{1}{2} H e^{2A} A_{\mu}(-q) \eta^{\mu\nu} \partial_r A_{\nu}(q)\right]
\,\right.\\
&&\left.
-\frac{1}{2}\left[iq^{\mu}A_{\mu}(-q)\partial_r\left(\frac{}{} H e^{2A} A_5(q)\right)\,+\,(q\leftrightarrow -q)\right]
\,\right.\\
&&\nonumber\left.
+\sum_{i=1,2}(-)^i\delta(r-r_i)\left[\frac{1}{2}i H e^{2A} q^{\mu}A_{\mu}(-q) A_5(q)\,+\,(q\leftrightarrow -q)\right]
\,\right.\\
&&\nonumber\left.
-\frac{1}{2}m^2 G e^{4A} A_5(-q)A_5(q)
-\frac{1}{2}\pi(-q)\partial_r\left[\frac{}{}- G e^{4A} \partial_r \pi(q)\right]
\,\right.\\
&&\nonumber\left.
+\sum_{i=1,2}(-)^i\delta(r-r_i)\left[-\frac{1}{2}G e^{4A} \pi(-q) \partial_r \pi(q) \right]
\,\right.\\
&&\nonumber\left.
-\frac{1}{2}\pi(-q) \partial_r\left[\frac{}{}-m G e^{4A} A_5(q)\right]
\,\right.\\
&&\nonumber\left.
-\frac{1}{2}A_5(-q) \left[\frac{}{}m G e^{4A} \partial_r \pi(q)\right]
\,\right.\\
&&\nonumber\left.
+\sum_{i=1,2}(-)^i\delta(r-r_i)\left[-\frac{1}{2}m G e^{4A} \pi(-q) A_5(q)\right]
\,\right.\\
&&\nonumber\left.
-\frac{1}{2}G e^{2A}\left[\frac{}{}q^2 \pi(-q)\pi(q)+m^2 \eta^{\mu\nu}A_{\mu}(-q)A_{\nu}(q)\right]
\,\right.\\
&&\nonumber\left.\, -\frac{1}{2}m G e^{2A} \,
\left[-iq_{\mu}\pi(-q)  \eta^{\mu\nu}A_{\nu}(q) \frac{}{}+ \, (q\rightarrow -q)\right]
\right\}\,.
\eeqs

Because of the presence of the boundaries, we also add generic boundary-localised kinetic terms for the vector in the form
\beqs
{\cal S}_D^{(1)} &=& \int \di^4 x \di r \sum_{i=1,2} (-)^i
\delta(r-r_i)\sqrt{-g_5}D_i \left\{-\frac{1}{4}\tilde{g}^{MN}\tilde{g}^{RS}F_{MR}F_{NS}\right\}\,\\
&=&\nonumber
\int \di^4 q \di r \sum_{i=1,2}(-)^i\delta(r-r_i)\left\{
-\frac{1}{2} D_i q^2 A_{\mu}(-q) P^{\mu\nu} A_{\nu}(q) \right\}\,,
\eeqs
and for the pseudo-scalar as in
{\small
\beqs
{\cal S}_C^{(1)}&=&\int \di^4 x \di r \sum_{i=1,2}(-)^i\delta(r-r_i)\sqrt{-g_5}\left\{
-\frac{1}{2} C_i \left[\frac{}{}\partial_{\mu}\pi+m A_{\mu}\right]
\tilde{g}^{\mu\nu} \left[\frac{}{}\partial_{\nu}\pi+ m A_{\nu}\right] \right\}\,\\
&=&\int \di^4 q \di r \sum_{i=1,2}(-)^i\delta(r-r_i)\left\{
-\frac{1}{2} C_i e^{2A}\left[\frac{}{}q_{\mu}\pi(-q)+i m A_{\mu}(-q)\right]\eta^{\mu\nu} 
\left[\frac{}{}q_{\nu}\pi(q)-i m A_{\nu}(q)\right] \right\}\,.\nonumber
\eeqs
}
The four constants $D_i$ and $C_i$  are  exhibited for completeness: they enter the process of holographic renormalisation, 
and we comment about them at the end of this Appendix.

The action contains mixing terms between the vector and pseudo-scalar. We hence add the following bulk gauge-fixing term 
\beqs
{\cal S}^{(1)}_{\xi}
&=&\int \di^4 q \di r \left\{ -\frac{H}{2\xi}
\left[q^{\mu}A_{\mu}(-q) + m\,i\frac{\xi}{H} G e^{2A} \pi(-q) + i \frac{\xi}{H} \partial_r\left(\frac{}{}H e^{2A} A_5(-q)\right)\right]\times\right.\nonumber\\
&&\left.\times
\left[q^{\nu}A_{\nu}(q) - m\,i\frac{\xi}{H} G e^{2A} \pi(q) - i \frac{\xi}{H} \partial_r\left(\frac{}{}H e^{2A} A_5(q)\right)\right]
\right\}\,, 
\eeqs
as well as the boundary-localised gauge fixing terms
{\small
\beqs
{\cal S}_M^{(1)}
&=&\int \di^4 q \di r \sum_{i=1,2}(-)^i\delta(r-r_i)\left\{
-\frac{1}{2M_i}\nonumber
\left[q^{\mu}A_{\mu}(-q) \frac{}{}- i M_i H e^{2A} A_5(-q) + i m M_i C_i e^{2A} \pi(-q) \right] \times\right.\\
&&\left.\times
\left[q^{\nu}A_{\nu}(q) \frac{}{}+ i M_i H e^{2A} A_5(q) - i m M_i C_i e^{2A} \pi(q) \right]\frac{}{}
\right\}\,.
\eeqs
}
The gauge-fixing parameter $\xi$ is in general a function of the radial direction $r$: because the fifth dimension is a segment, the $U(1)$ in five dimensions gives rise to an infinite tower of $U(1)$ gauge theories in four dimensions, each of which is spontaneously broken, and each of which could in principle be gauge-fixed independently. For simplicity, we set $\xi$ to a constant. The boundary-localised (dimensionful) gauge-fixing parameters $M_i$ are independent of the bulk dynamics, and again their arbitrariness corresponds to the arbitrariness in gauge fixing the boundary $U(1)$. We make the choice $M_i=\frac{\xi}{D_i}$, so that the action of the longitudinally polarised part of the vectors can be obtained from the transverse one by replacing $q^2\rightarrow q^2/\xi$, as in the $D=4$ case discussed in Appendix~\ref{Sec:A1}. We find the  action of the spin-1 fields to vanish on-shell when imposing the equations of motion and boundary conditions, which we list as follows:
\beqs
\left[\frac{}{} q^2 H  -\partial_r \left( H e^{2A} \partial_r \right) + m^2 G e^{2A}\right] P^{\mu\nu}A_{\mu}(q,r)&=&0\,,\\
\label{Eq:bcU15D}
\left.\left[\frac{}{}H e^{2A} \partial_r +q^2 D_i + m^2 C_i e^{2A} \right] P^{\mu\nu}A_{\nu}(q,r)\right|_{r=r_i}&=&0\,,\\
\left[\frac{}{} \frac{q^2}{\xi} H  -\partial_r \left( H e^{2A} \partial_r \right) + m^2 G e^{2A}\right] \frac{q^{\mu}q^{\nu}}{q^2}A_{\mu}(q,r)&=&0\,,\\
\left.\left[\frac{}{}H e^{2A} \partial_r +\frac{q^2}{\xi} D_i + m^2 C_i e^{2A} \right] \frac{q^{\mu}q^{\nu}}{q^2}A_{\nu}(q,r)\right|_{r=r_i}&=&0\,.
\eeqs

The equations for the (pseudo-)scalars $A_5$ and $\pi$ look significantly more complicated, until one exploits gauge-invariance by introducing the following re-definitions:
\beqs
A_5&\equiv&\frac{1}{m}\left(\frac{m X}{e^{4A} G} - \partial_r \pi\right)\,,\\
\pi&\equiv&Y + \frac{m \partial_r X}{q^2 e^{2A} G}\,.
\eeqs
The equations  for  the physical (gauge invariant) scalar field $X$ decouples from $Y$. The two obey the following equations of motion and boundary conditions:
\beqs
\left[\frac{}{}\partial_r^2  +\left(-2 \partial_r A -\frac{\partial_r G}{G}\right)\partial_r  + \left( -e^{-2A} q^2 -\frac{m^2 G}{H}\right)\right]X(q,r)&=&0\,,\\
\label{Eq:bcXU15D}
\left.\left[\frac{}{}C_i \partial_r  \,+\, G\,\right] X(q,r) \right|_{r=r_i}&=&0\,,\\
\left[\frac{}{}\partial_r^2  +\left(2 \partial_r A +\frac{\partial_r H}{H}\right)\partial_r  
+ \left( -e^{-2A} \frac{q^2}{\xi} -\frac{m^2 G}{H}\right)\right]Y(q,r)&=&0\,,\\
\left.\left[\frac{}{}He^{2A}\partial_r  \,+\,\left( \frac{D_i}{\xi} q^2 +\frac{}{}  m^2 C_ie^{2A}\right)\right] Y(q,r)\right|_{r=r_i}&=&0\,.
\eeqs 
The equations for the gauge-dependent $Y$ are identical (up to an inconsequential multiplicative factor) to those obeyed by the longitudinal polarisations of the vectors $\frac{q^{\mu}q^{\nu}}{q^2}A_{\nu}(q,r)$.

In this paper we are interested only in computing the physical spectrum of states appearing as isolated poles in the 2-point functions involving operators of the dual field theory. This can be obtained by taking functional derivatives of the bulk action 
evaluated on-shell, with respect to properly defined (and properly normalised) boundary-localised sources (see for instance~\cite{Skenderis:2002wp,Papadimitriou:2004ap}). In doing so, one comes to realise that for asymptotically-AdS backgrounds the divergences can be cancelled by the counter-terms $D_i$ and $C_i$.

The procedure we follow is superficially very different, but in fact yields the same results. By imposing the IR and UV boundary conditions on the differential equations, and hence over-constraining the system, one finds a discrete set of $q^2$ corresponding to the zeros, rather than the poles, of the relevant correlation functions, and hence the process seems to differ from the one of physical interest by a Legendre transform. Yet, because   we are  setting $C_i=0=D_i$,  and hence keeping divergent additive contributions (polynomial in $q^2$) to the 2-point functions, by solving the equations for finite regulators $r_i$, and afterwards taking the physical limits and removing the regulators, the results we obtain for the spectrum exactly coincide with the isolated poles of the physical correlator.

\subsection{2-forms in $D=5$ dimensions}
\label{Sec:B3}

The discussion in Appendix~\ref{Sec:B2} generalises non-trivially to higher-order $p$-forms. For a 2-form $B_{MN}$ one defines the field-strength 
\beqs
G_{MNT}&=&3\partial_{[M}B_{NT]}\,=\,\partial_MB_{NT}+\partial_NB_{TM}+\partial_TB_{MN}\,,
\eeqs
having made use of the anti-symmetry of $B_{MN}$. Under the gauge transformation 
\beqs
B_{MN}&\rightarrow B_{MN}-{2}{}\partial_{[M}\alpha_{N]}\,,
\label{Eq:U1form}
\eeqs
with $\alpha_M$ a vector depending on the coordinates, $G_{MNT}$ is invariant. One then proceeds in a similar way as in the case of the $1$-form and $0$-form in Appendix~\ref{Sec:B2}: by introducing a $1$-form transforming with a shift under the transformation in~(\ref{Eq:U1form})
\beqs
A_M&\rightarrow & A_M+m \alpha_M\,,
\eeqs
with $m$ a constant, one finds that a gauge-invariant 2-form is given by
\beqs
{\cal H}_{MN}&=&F_{MN}+ m B_{MN}\,,
\eeqs
where $F_{MN}=2\partial_{[M}A_{N]}$. This procedure generalises the Higgs mechanism to $p$-forms in $D$ dimensions. The process that eventually leads to the equations of motion for the fluctuations, boundary conditions, and (four-dimensional) spectrum of physical states mimics what is done for $1$-forms in Appendix~\ref{Sec:B2}. We report it in detail, highlighting some important subtleties.

The action to be added to the sigma-model coupled to gravity takes the form:
\beqs
{\cal S}_5^{(2)} &=&\int \di^4x\di r \sqrt{-g_5}\left\{ -\frac{1}{4}H\,g^{MR} g^{NS}\,{\cal H}_{MN} {\cal H}_{RS}\right.+\\
&&\nonumber\left.-\frac{1}{12}K\, g^{MR}\,g^{NS}\,g^{TU} G_{MNT}G_{RSU}\right\}\,,
\label{Eq:2form}
\eeqs
where $H$ and $K$ are functions of the background values of the sigma-model scalars $\Phi^a$.

After Fourier-transforming all the fields, the action can be written  as follows.
\beqs
{\cal S}_5^{(2)} &=&\int \di^4q\di r \left\{\nonumber
-\frac{1}{2} H e^{2A}\left[\frac{}{}\partial_rA_{\mu}(-q)
+ m B_{5\mu}(-q)\right]\eta^{\mu\nu}\left[\frac{}{}\partial_rA_{\nu}(q)+ m B_{5\nu}(q)\right]\right.\\
&&
\left.
-\frac{1}{2}H q^2 e^{2A} A_5(-q)A_5(q)
\right.\nonumber\\
&&\nonumber\left.
-\frac{1}{2}H e^{2A}\left[\frac{}{}i A_5(-q)\left(\frac{}{}q^{\mu}\partial_rA_{\mu}(q)+m q^{\mu}B_{5\mu}(q)\right)\,+\,(q\leftrightarrow -q)\right]
\right.\\
&&\left.
-\frac{1}{2}H A_{\mu}(-q)\,q^2 P^{\mu\nu} A_{\nu}(q)
\right.\\
&&\nonumber\left.
-\frac{1}{4}H m^2 B_{\mu\nu}(-q) \,\eta^{\mu\rho}\eta^{\nu\sigma}\, B_{\rho\sigma}(q)
\right.\\
&&\nonumber\left.
-\frac{1}{2}H\eta^{\mu\nu}\left[\frac{}{}i m q^{\rho}B_{\rho\mu}(-q)A_{\nu}(q)\,+\,(q\leftrightarrow -q)\right]
\right.\\
&&\nonumber\left.
-\frac{1}{4} B_{\mu\nu}(-q)\,\eta^{\mu\rho}\eta^{\nu\sigma}\left[\frac{}{}-\partial_r\left(\frac{}{}K \partial_rB_{\rho\sigma}(q)\right)\right]
\right.\\
&&\nonumber\left.
+\sum_{i=1,2}(-)^i\delta(r-r_i)\left[-\frac{1}{4}K B_{\mu\nu}(-q)\,\eta^{\mu\rho}\eta^{\nu\sigma}\,\partial_rB_{\rho\sigma}(q)\right]
\right.\\
&&\nonumber\left.
-\frac{1}{2}K \,B_{5\mu}(-q)\,q^2P^{\mu\nu}\,B_{5\nu}(q)
\right.\\
&&\nonumber\left.
-\frac{1}{2} K\eta^{\mu\nu}\left[\frac{}{}-i q^{\rho}\partial_rB_{\rho\mu}(-q)\, \,B_{5\nu}(q)\,+\,(q\leftrightarrow -q)\right]
\right.\\
&&\nonumber\left.
-\frac{1}{4}K e^{-2A} B_{\mu\nu}(-q)\,q^2 \,P^{\mu\rho} P^{\nu\sigma}\,B_{\rho\sigma}(q)
\frac{}{}\right\}\,.
\eeqs
Notice that $B_{\mu\nu}(-q)\, \,P^{\mu\rho} P^{\nu\sigma}\,B_{\rho\sigma}(q)
= B_{\mu\nu}(-q)\, \,\left(\eta^{\mu\rho} \eta^{\nu\sigma}-2 \frac{q^{\mu}q^{\rho}}{q^2}\eta^{\nu\sigma}\right)\,B_{\rho\sigma}(q)$.

Besides the bulk action, we also add boundary-localised kinetic terms:
{\small
\beqs
{\cal S}_{E}^{(2)}
&=&\int \di^4x\di r \sum_{i=1,2}(-)^i \delta(r-r_i)\sqrt{-g_5}\left\{
-\frac{1}{12}E_i K \tilde{g}^{\mu\sigma}\tilde{g}^{\nu\tau}\tilde{g}^{\rho\omega}G_{\mu\nu\rho}G_{\sigma\tau\omega}
\right\}\,,\\
&=&\int \di^4q\di r \sum_{i=1,2}(-)^i \delta(r-r_i)\left\{
-\frac{1}{4}B_{\mu\nu}(-q)\,\left(\eta^{\mu\rho}\eta^{\nu\sigma}-2\frac{q^{\mu}q^{\rho}}{q^2}\eta^{\nu\sigma}\right)\,
e^{-2A}K\,E_i q^2\,B_{\rho\sigma}(q)\nonumber
\right\}\,,\\
{\cal S}_{D}^{(2)}&=&\int \di^4x\di r \sum_{i=1,2}(-)^i \delta(r-r_i)\sqrt{-g_5}\left\{
-\frac{1}{4}D_i H  \tilde{g}^{\mu\sigma}\tilde{g}^{\nu\tau} {\cal H}_{\mu\nu}  {\cal H}_{\sigma\tau}
\right\}\\
\nonumber
&=&\int \di^4q\di r \sum_{i=1,2}(-)^i \delta(r-r_i)\left\{
-\frac{1}{4} D_i H 
\left[\frac{}{}q_{\mu}A_{\nu}(-q)-q_{\nu}A_{\mu}(-q)+ i m B_{\mu\nu}(-q)
\right]
\eta^{\mu\rho}\eta^{\nu\sigma}\times\right.\nonumber\\
&&\left.
\times\left[\frac{}{}q_{\rho}A_{\sigma}(q)-q_{\sigma}A_{\rho}(q)- i m B_{\rho\sigma}(q)
\right]\nonumber
\right\}\,.
\eeqs
}
The parameters $E_i$ and $D_i$ play the analogous role of the boundary-localised counter-terms introduced when dealing with
1-forms in Appendix~\ref{Sec:B2}.

The decomposition in four-dimensional language of the original fields leads to $A_5$ behaving as a pseudo-scalar, $A_{\mu}$ and $B_{5\mu}$ behaving as vectors and $B_{\mu\nu}$ being a 2-form. We want to eliminate mixing terms between forms of different orders, by adding bulk and boundary  gauge-fixing terms:
{\small
\beqs
{\cal S}_{\Xi\,,2}^{(2)}
&=&\int \di^4q\di r \left\{-\frac{e^{2A} K}{2\Xi}\nonumber
\left[\frac{}{}e^{-2A}q^{\rho} B_{\rho\mu}(-q)+i\frac{\Xi}{K}  \partial_r \left(K\frac{}{}B_{5\mu}(-q)\right)
+i\frac{\Xi}{K} m H A_{\mu}(-q)\right]
\times\right.\nonumber\\
&&\left.\frac{}{}\times  \eta^{\mu\nu} 
\left[e^{-2A}\frac{}{}q^{\sigma} B_{\sigma\nu}(q)-i\frac{\Xi}{K}  \partial_r \left(K\frac{}{}B_{5\nu}(q)\right)
-i\frac{\Xi}{K} m H A_{\nu}(q)\right]
\,\right\}\,,\\
{\cal S}_{N\,,2}^{(2)}\nonumber
&=&\int \di^4q\di r \sum_{i=1,2}(-)^i \delta(r-r_i)\left\{
-\frac{K e^{2A}}{2N_i}\eta^{\mu\nu}
\times\right.\nonumber\\
&&\times
\left[\frac{}{}e^{-2A}q^{\rho}B_{\rho\mu}(-q)-iN_iB_{5\mu}(-q)+i m \frac{D_i H N_i}{K} A_{\mu}(-q)\right]\times\\
&&\left.\frac{}{}\times \nonumber
\left[\frac{}{}e^{-2A}q^{\sigma}B_{\sigma\nu}(q)+iN_iB_{5\nu}(q)-i m \frac{D_i H N_i}{K} A_{\nu}(q)\right]\right\}\,,\\
{\cal S}^{(2)}_{\xi,1}
&=&\int \di^4q \di r\left\{\nonumber
-\frac{K}{2\xi}\left[q^{\mu}B_{5\mu}(-q)-i\xi m e^{2A}\frac{H}{K} A_{5}(-q)\right]
\left[q^{\nu}B_{5\nu}(q)+i\xi m e^{2A}\frac{H}{K} A_{5}(q)\right]+\right.\\
&&\left.-\frac{H}{2\xi}\left[q^{\mu}A_{\mu}(-q)+i\frac{\xi}{H}\partial_r\left(\frac{}{}He^{2A}A_5(-q)\right)\right]
\left[q^{\nu}A_{\nu}(q)-i\frac{\xi}{H}\partial_r\left(\frac{}{}He^{2A}A_5(q)\right)\right]
\right\},\\
{\cal S}^{(2)}_{M,1}
&=&\int \di^4q \di r \sum_{i=1,2}(-)^i\delta(r-r_i)
\left\{
-\frac{H}{2M_i}\left[\frac{}{} q^{\mu}A_{\mu}(-q) - i e^{2A}M_i A_5(-q)\right]\times\right.\\
&&\left.\times
\left[\frac{}{} q^{\nu}A_{\nu}(q) + i e^{2A}M_i A_5(q)\right]\nonumber
\right\}\,.
\eeqs
}

The first two such terms decouple the 2-form from lower-order forms, by exploiting the vectorial part of the gauge invariance (the transformations controlled by the $\alpha_{\mu}$ parameter). The parameter $\Xi$ is a generic function of $r$, but for simplicity we choose it to be a constant, while we fix the boundary-localised $N_i$ to obey the relation $N_i=\Xi/E_i$. There is an additional residual gauge symmetry, that allows one to remove mixing of the vectors $B_{5\mu}$ and $A_{\mu}$ with the pseudo-scalar $A_5$ by adding the last two gauge-fixing terms controlled by $\xi$ and $M_i$.

The final result of the exercise for the 2-forms is that the bulk equations and boundary conditions for the transverse polarisations read
\beqs
\left[\frac{}{}Kq^2 e^{-2A}-\partial_r\left(\frac{}{}K \partial_r\right)+H m^2\right] P^{\mu\rho} P^{\nu\sigma} B_{\rho\sigma}(q,r)&=&0\,,\\
\left.\frac{}{}
\,
\,\left[K\,E_i q^2 e^{-2A}+ K \partial_r +D_i H m^2\frac{}{}\right] P^{\mu\rho} P^{\nu\sigma} B_{\rho\sigma}(q,r)\right|_{r=r_i} &=&0\,,\label{Eq:bc2form5d}
\eeqs
while the longitudinal components obey  equations obtained by the replacement  $q^2\rightarrow q^2/\Xi$.

For the transverse polarisations of $A_{\mu}$ and $B_{5\mu}$, we define the generalised $U(1)$ gauge invariant field $X_{\mu}$ via the relation
\beqs
B_{5\mu}&\equiv& \frac{1}{m}\left(\frac{m X_{\mu}}{e^{2A} H} - \partial_r A_{\mu}\right)\,,
\eeqs
and its complementary (transverse) field $Y_{\mu}$ via
\beqs
P_\mu{}^\nu A_\nu&\equiv&Y_{\mu}+\frac{m \partial_r X_{\mu}}{q^2  H}\,.
\eeqs
By making use of the equations of motion for $A_{\mu}$, the equations for the physical vector $X_{\mu}$ decouple and hence read:
\beqs
\left[\frac{}{}\partial_r^2 \,-\,
\frac{\partial_r H}{H}
   \partial_r
+\left(-e^{-2 A} q^2 - m^2 \frac{H}{K} \right) 
   \right] X_{\mu}(q,r)&=&0\,,
\eeqs
subject to the boundary conditions
\beqs
\left.\frac{}{}\left[ \partial_r  +\frac{1}{D_i} \right] X_{\mu}(q,r)\right|_{r=r_i}&=&0\,.
\label{Eq:bc1form5D}
\eeqs
The equations for $Y_{\mu}$ are then
\beqs
\left[\frac{}{}\partial_r^2 \,+\,
\frac{\partial_r K}{K}
   \partial_r
+\left(-e^{-2 A} \frac{q^2}{\Xi} - m^2 \frac{H}{K} \right) 
   \right]Y_{\mu}(q,r)&=&0\,,
\eeqs
subject to the boundary conditions
\beqs
\left.\frac{}{}\left[ \partial_r  +\left(\frac{e^{-2A} q^2}{N_i} + \frac{D_i m^2 H}{K}\right)\right] Y_{\mu}(q,r)\right|_{r=r_i}&=&0\,.
\eeqs
By choosing $N_i=\Xi/E_i$ we see that the equations and boundary conditions for $Y_\mu$ explicitly depend on the generalised $U(1)$ gauge-fixing parameter choice $\Xi$, and furthermore that the bulk equations and boundary conditions for $Y_{\nu}$ are identical to those of the transverse $P^{\mu\rho}P^{\nu\sigma}B_{\rho\sigma}$, up to the replacement $q^2\rightarrow \frac{q^2}{\Xi}$. This is the analogue of what we found in the case of a spontaneously broken ordinary $U(1)$ in Appendix~\ref{Sec:B2}: 
the transverse components of the vector $Y_{\mu}$ are Higgsed into $B_{\mu\nu}$ and provide it with the 2 additional polarisations that turn it from a massless 2-form (dual to a scalar, with 1 d.o.f.) into a massive 2-form (dual to a massive vector, with 3 d.o.f.).

In order to decouple the equations of the longitudinal polarisations of the vectors, we slightly modify the definition of $X^L_{\mu}$ and $Y^{L}_{\mu}$ according to
\beqs
B_{5\mu}^L&\equiv& \frac{1}{m}\left(\frac{m X_{\mu}^L}{e^{2A} H} - \partial_r A_{\mu}^L\right)\,,\\
A_{\mu}^L&\equiv&Y_{\mu}^L+\xi\frac{m \partial_r X_{\mu}^L}{q^2  H}\,,
\eeqs
where the suffix $^L$ indicates the component projected along $q^{\mu}$. The final equations and boundary conditions read as follows:
\beqs
0&=&\left[\frac{}{}\partial_r^2 \,-\,
\frac{\partial_r H}{H}
   \partial_r 
+\left(-e^{-2 A} \frac{q^2}{\xi} - m^2 \frac{H}{K} \right) 
   \right] X_{\mu}^L(q,r)\,,\\
0&=&\left.\frac{}{}\left[ \partial_r +\frac{1}{D_i} \right]
X_{\nu}^L(q,r)\right|_{r=r_i}\,,\\
0&=&\left[\frac{}{}\partial_r^2 \,+\,
\frac{\partial_r K}{K}
   \partial_r   
+\left(-e^{-2 A} \frac{q^2}{\xi \Xi} - m^2 \frac{H}{K} \right) 
   \right] Y_{\mu}^L(q,r)\,,\\
0&=&\left.\frac{}{}\left[ \partial_r  +\left(\frac{e^{-2A} q^2}{\xi N_i} + \frac{D_i m^2 H}{K}\right)\right]
Y_{\mu}^L(q,r)\right|_{r=r_i}\,,
\eeqs
where we have made use of the replacement $M_i=\frac{\xi}{D_i}$, thanks to which these equations are identical to those of the transverse polarisations, except for the replacement $q^2\rightarrow \frac{q^2}{\xi}$. In particular, this confirms that none of the longitudinally-polarised vector fields is physical.

Finally, the scalar sector contains only $A_5$, and has been decoupled from all other fields. The bulk equation is
\beqs
0&=&\left(\frac{q^2}{\xi} + m^2 e^{2A} \frac{H}{K}\right)A_5(q) - \partial_r \left[ \frac{1}{H}\partial_r \left(\frac{}{}H e^{2A} A_5(q)\right)\right]\,,
\eeqs
subject to the boundary conditions
\beqs
0&=&\left.\frac{1}{D_i} A_5(q)+\frac{1}{H e^{2A}} \partial_r \left[\frac{}{}H e^{2A}A_5(q)\right]\right|_{r=r_i}\,,
\eeqs
where once more we have chosen $M_i=\frac{\xi}{D_i}$. We conclude by observing that the scalar $He^{2A} A_5$ obeys identical equations of motion and boundary conditions as $q^{\nu}X^L_{\nu}$, as expected.

In summary, the physical masses can be computed by looking at the transverse polarisation of the 2-form $B_{\mu\nu}$, and at the gauge-invariant combination $X_{\mu}$ of the transverse polarisations of $A_{\mu}$ and $B_{5\mu}$. All other fields---the longitudinal polarisations of $B_{\mu\nu}$, the gauge dependent combination $Y_{\mu}$ of the transverse polarisations of the vectors, both of the $X^{L}_{\mu}$ and $Y^{L}_{\mu}$ combinations of the longitudinal polarisations of the vectors, and the pseudo-scalar $A_5$---are unphysical and gauge-dependent remnants of the Higgs mechanism in the generic $R_{\xi}$ gauge.

\end{document}